\pdfoutput=1
\documentclass[journal]{IEEEtran}
\usepackage{graphicx}
\usepackage{subfigure}
\usepackage{multirow}
\usepackage{array}
\usepackage{footnote}

\usepackage[ruled,linesnumbered]{algorithm2e}
\usepackage{algorithmic}
\usepackage{dsfont}
\usepackage{amsfonts,amsmath}
\usepackage{url,cite}
\usepackage{color}
\usepackage{bm}
\usepackage[table,xcdraw]{xcolor}

\usepackage{amssymb}

\usepackage{ntheorem}
\newtheorem*{remark}{Remark}

\def\obsset{\boldsymbol{\mathsf{Y}}}

\begin{document}

\title{Guided Deep Generative Model-based Spatial Regularization for Multiband Imaging Inverse Problems}

\author{Min~Zhao,~\IEEEmembership{Student Member,~IEEE,}
        Nicolas Dobigeon,~\IEEEmembership{Senior Member,~IEEE}
       and Jie~Chen,~\IEEEmembership{Senior Member,~IEEE}

\thanks{M. Zhao and J. Chen are with School of Marine Science and Technology,
Northwestern Polytechnical University, Xi'an 710072, China (e-mail:
minzhao@mail.nwpu.edu.cn; dr.jie.chen@ieee.org).}
\thanks{Nicolas Dobigeon is with University of Toulouse, IRIT/INPENSEEIHT,
CNRS, 2 rue Charles Camichel, BP 7122, 31071 Toulouse Cedex
7, France and also with the Institut Universitaire de France
(IUF), France (e-mail: Nicolas.Dobigeon@enseeiht.fr).}}


\maketitle

\begin{abstract}
When adopting a model-based formulation, solving inverse problems encountered in multiband imaging requires to define spatial and spectral regularizations. In most of the works of the literature, spectral information is extracted from the observations directly to derive data-driven spectral priors. Conversely, the choice of the spatial regularization often boils down to the use of conventional penalizations (e.g., total variation) promoting expected features of the reconstructed image (e.g., piecewise constant). In this work, we propose a generic framework able to capitalize on an auxiliary acquisition of high spatial resolution to derive tailored data-driven spatial regularizations. This approach leverages on the ability of deep learning to extract high level features. More precisely, the regularization is conceived as a deep generative network able to encode spatial semantic features contained in this auxiliary image of high spatial resolution. To illustrate the versatility of this approach, it is instantiated to conduct two particular tasks, namely multiband image fusion and multiband image inpainting. Experimental results obtained on these two tasks demonstrate the benefit of this class of informed regularizations when compared to more conventional ones.
\end{abstract}


\IEEEpeerreviewmaketitle

\section{Introduction}
\label{sec:intro} \IEEEPARstart{M}{ultiband} imaging consists in acquiring multivariate images whose pixels are associated with a vector of measurements. Conventional red-green-blue (RGB) imaging is the simplest instance of this technique where a $3$-dimensional vector encodes the reflectance measured in each spatial location at three specific wavelengths. This principle can be generalized by designing elaborated dedicated optical systems, leading to multispectral imaging~\cite{demars2015multispectral} or hyperspectral imaging~\cite{landgrebe2002hyperspectral}. Such imaging techniques have been widely used in remote sensing, not only for earth observation with applications such as military surveillance, environmental monitoring and disaster assessment \cite{ghamisi2017advances} but also planetology \cite{doute2005nature}. Beyond its ubiquitous use to sense the electromagnetic spectrum in the visible range, multiband imaging has shown to be of great value to sound the universe where crucial information is located in the near-infrared, e.g., with the recently launched James Webb spatial Telescope (JWST) \cite{guilloteau2020hyperspectral, guilloteau2020simulated}. In the contexts of astronomy and astrophysics, multiband images can  also be acquired by integral field spectrographs, such as the multi-unit spectroscopic explorer (MUSE) operated on the Very Large Telescope (VLT)~\cite{soulez2011restoration}. At radically different scales, multi-dimensional data cubes can be also acquired by scanning transmission electron microscopy (STEM) \cite{monier2018reconstruction}, magnetic resonance spectroscopy (MRS) \cite{basic1996nuclear} or dynamic positron emission tomography (PET)~\cite{cavalcanti2018unmixing}. With such imaging modalities, the ``spectra'' collected in each spatial position (pixels or voxels) do not come from the sensing of the electromagnetic spectrum but is associated with a physical process such as an electron energy loss (STEM-EELS), a magnetic field around atomic nuclei (MRS) or the temporal response of a radiotracer (dynamic PET).

However, due to an insurmountable yet intrinsic spatial vs. spectral trade-off, multiband images are characterized by a limited spatial resolution, much lower than the more conventional imaging techniques. Depending on the considered applicative contexts, these images can be affected by blurs, undersampling or high level noises. Part of the measurements may also unavailable when the acquisition of the full data cube is impossible. This can arise in several applicative scenarios due to constraints imposed by the instrumental process or the measurement protocol, for instance to reduce the sample damages of sensitive materials. These shortcomings may significantly impair the exploitability of the images and the subsequent analyses, such as target detection, material identification and classification. To cope with these limitations, one strategy consists in solving an inverse problem which aims at recovering a full representation of higher quality from the degraded measurements. Some archetypal examples of such inversion tasks specifically dedicated to multiband imaging include denoising, deconvolution, inpainting, single image superresolution and  multiple image fusion.

Numerous works were devoted to the design of multiband imaging inversion techniques. They can be primarily divided into conventional model-based and more recent learning-based methods. The former usually solves the multiband imaging inverse problems by prescribing a certain regularization able to embed expected characteristics of the restored image. This strategy has shown to be particularly appealing to capture the spectral redundancy of the images, e.g., by imposing a low rank structure. This structure can be further inferred and/or informed by analyzing the spectral content of the acquired multiband image itself, e.g., by conducting a principal component analysis~\cite{wei2015hyperspectral, dian2020regularizing}. Regarding the spatial regularizations, numerous handcrafted model-based priors have been proposed, such as total variation~\cite{simoes2014convex}, sparsity~\cite{xue2021spatial}, low-rankness~\cite{chang2020weighted} and dictionary learning~\cite{wei2015hyperspectral}. However, selecting an appropriate regularizer to match the intrinsic properties of the image is a nontrivial task. More importantly, these models can hardly incorporate the richness of the spatial content of the images. Devising sophisticated but sufficiently generic spatial regularizations able to capture the diversity of the image properties is generally accompanied by a significant increase of the resulting computational burden.

Conversely learning-based methods have been recently proposed to circumvent this bottleneck and have became a hotspot thanks to its superior capability to excavate high-level features. This kind of methods aim at learning a nonlinear mapping from the raw measurements to the restored image where the image priors are implicitly encoded in the network parameters. However, such approaches can be cast as black-box techniques and generally lacks of physical interpretability.

To bridge the gap between the conventional model-based and learning-based methods, this paper introduces a new and smart framework to spatially regularize multiband imaging inverse problems. In particular we capitalize on appealing scenarios when the imaging protocol provides an image of the same observed scene at a higher spatial resolution. Such scenarios arise naturally in various applicative contexts. For instance, this is the case when images of complementary spatial and spectral resolutions have to be fused or when the acquisition of a multiband image to be restored is concomitant with the acquisition of an auxiliary image of higher spatial resolution. The proposed method incorporates a learning-based spatial regularization into a conventional model-based formulation. This regularization consists of a pretrained deep generative decoder informed by the auxiliary image of high spatial resolution. The main contributions of this work can be summarized as follows:
\begin{itemize}
    \item We propose a new way to regularize multiband imaging inverse problems by means of a deep generative network able to encode  prior information learnt from a high spatial resolution complementary image. As an example, this generative network is chosen as a guided deep decoder (GDD) recently proposed in the literature. However, the proposed framework may not be limited to this particular choice and may remain valid for any guided generative model that would be designed in future works. The relevance of this informed regularization is illustrated by comparing experimental results with those obtained when the generative model is chosen as a variational encoder (VAE) not trained on the auxiliary image specifically.
  \item Instead of resorting to an automatic differentiation technique to minimize the resulting optimization criterion, we devise a splitting-based strategy which has the great advantages of decomposing the initial problem into simpler subproblems. In particular for some subproblems, closed-form algorithmic updates can be implemented. We empirically validate this choice by comparing the restoration results reached by the proposed strategy with those obtained by the Adam optimizer.
  \item To illustrate the versatility of the proposed framework, it is instantiated for two ubiquitous inversion tasks, namely image fusion and image inpainting. Extensive simulation results obtained from experiments conducted for these two tasks show that the proposed framework competes favorably with respect to state-of-the-art inversion methods.
\end{itemize}

The reminder of this paper is organized as follows. Recent works about multiband imaging inversion are reviewed in Section~\ref{sec:related_work}, with a particular focus on the design of model-based and learning-based regularizations. The problem addressed in this paper is stated in Section~\ref{sec:problem}. Section~\ref{sec:proposed} introduces the proposed generic framework to perform multiband imaging inversion. This framework is instantiated in Section~\ref{sec:appli} for two particular yet ubiquitous tasks, namely fusion and inpainting. Section~\ref{sec:experiment} reports extensive experimental results obtained for these two tasks. Section~\ref{sec:conclusion} concludes the paper.

\section{Related Works}
\label{sec:related_work}
\subsection{Model-based regularizations}
A significant amount of regularizations have been designed to describe the underlying image characteristics and to improve the inversion task. Reviewing the whole literature would be a titanic task and is out-of-the-scope of this paper. Only a few directions are listed in the sequel of this section. Numerous works such as~\cite{simoes2014convex,chang2015anisotropic} and \cite{wang2017hyperspectral} use a conventional total variation (TV) to preserve edges and promote piece-wise constant content. The methods~\cite{xue2021spatial,wei2015hyperspectral} impose  low-rank structures of the multiband image, e.g., by decomposing the data into a set of basis elements or dictionary. The representation coefficients are then associated to sparsity-promoting penalizations, such as the $\ell_0$-pseudo-norm or the $\ell_1$-norm. The intrinsic property of image (non-local) self-similarity is another key strategy to exploit the redundancy arising in multiband images. The work~\cite{xue2021spatial} proposes a  sparse low-rank representation model to perform multiband image super-resolution. In~\cite{wang2016adaptive}, a 3D nonlocal sparse representation is introduced to take advantage of non-local similarity in both spatial and spectral domains.
Another strategy consists in resorting to a superpixel segmentation step to group spectrally similar pixels. The work~\cite{fan2021hyperspectral} exploits the entropy rate superpixel segmentation method to divide the image into superpixels that are subsequently processed to ensure spectral smoothness and to preserve image details.

\subsection{End-to-end learning-based methods}
To avoid designing hand-craft model-based priors, one alternative consists in formulating the inversion problem as a learning task by leveraging on the generalization ability of deep neural networks. It is then expected that the spatial and spectral redundancies intrinsic to the images are learnt from the training data set to be subsequently used as an implicit prior while solving the inversion problem.
In~\cite{dong2019deep}, the authors design a deep convolutional neural network (CNN) for hyperspectral image restoration, which uses a modified U-net and decomposes the 3D filtering into 2D spatial filtering and 1D spectral filtering to reduce the number of parameters. In~\cite{wong2020hsi}, a channel attention mechanism is used to capture spectral correlation information, and a local discriminative network is proposed to exploit a certain spatial continuity. The authors in~\cite{liu2020psgan} introduce a generative adversarial network to perform a pan-sharpening task. In~\cite{qian2021hyperspectral}, a self-supervised algorithm is proposed for hyperspectral image restoration. A 2D image denoiser pretrained on gray or red-green-blue (RGB) images is used as a backbone model and then fine-tuned on hyperspectral image band-by-band.  The work~\cite{zhang2020deep} proposes an unsupervised deep framework for hyperspectral super-resolution. In~\cite{sidorov2019deep}, a deep hyperspectral prior algorithm is designed for hyperspectral restoration. It is based on 3D convolutional networks able to jointly learn the spatial and local spectral information.

\subsection{Embedding learnt image prior into model-based methods}
Deep architectures are known to demonstrate a certain superiority in extracting image properties efficiently. However end-to-end deep learning methods generally lack from explicability and, more importantly, do not capitalize on the knowledge about the acquisition process. Besides using handcrafted priors, a new trend consists in embedding an prior knowledge learnt by the deep networks into more conventional model-based iterative optimization algorithm. The works~\cite{dian2018deep,wang2021hyperspectral,lin2021admm} exploit the output of deep networks as a constrained term to regularize the optimization problem. This regularizer term is chosen simple and convex, such as the squared Euclidean distance between the trained solution and the target estimation, which avoids heavy computation. Plug-and-play priors have also received a great attention in the context of multiband imaging inversion. The state-of-the-art denoising algorithms, such as CNN-based denoisers, are usually plugged into this framework as the proximity operator to capture the instinct spatial structures of images. For instance, the works~\cite{dian2020regularizing,lai2022deep} decompose the optimization problem into iterative subproblems. Specifically, one
of the subproblem can be cast as a proximal mapping related to the image prior model. Based on this interpretation, this subproblem can then be solved using a deep denoising operator, which incorporates deep priors into the estimation. Unrolling state-of-the-art optimization-based algorithms is another route followed by several recent works. It unfolds iterative optimization algorithms to derive a counterpart on the form of a trainable deep architectures. It allows the involved parameters to be learnt with the restored image jointly. For example, the work~\cite{ramirez2021ladmm} unrolls  alternating direction methods of multipliers (ADMM) as deep networks for hyperspectral super-resolution task, and the work~\cite{xie2020mhf} unfolds the  proximal gradient algorithm into deep networks for multiband image fusion.

\section{Problem statement}
\label{sec:problem}

This work considers a set $\obsset=\left\{\mathbf{Y}_1,\ldots,\mathbf{Y}_K\right\}$ of $K$ acquired multiband images $\mathbf{Y}_k \in \mathbb{R}^{B_k \times N_k}$ ($k=1,\ldots,K$) where $B_k\geq 1$ and $N_k\geq 1$ denote the numbers of bands (or channels) and pixels, respectively. These observations are assumed to be related to an unknown (latent) image $\mathbf{X} \in \mathbb{R}^{B\times N}$ through the direct model
\begin{equation}\label{eq:forwardmodel_generic}
    \mathcal{H}\left(\obsset\right) = \mathcal{M}\left(\mathbf{X}\right) + \mathbf{N}
\end{equation}
where $\mathcal{M}\left(\cdot\right)$ represents the forward operators mapping from the latent space to the observation space. In this work, the operator underlying $\mathcal{M}\left(\cdot\right)$ is assumed to be perfectly known and can describe various spatial or spectral degradations including spatial blurring, regular or irregular spatial subsampling, spectral filtering, etc. The operator $\mathcal{H}(\cdot)$ aims at selecting and rearranging the data from the set $\obsset$ to form the observations as provided by the sensors. For instance, it may select one multiband image from the $K$ acquired images $\mathbf{Y}_1,\ldots,\mathbf{Y}_K$ when this unique image is intended to be a spatially and/or spectrally degraded version of the latent image $\mathbf{X}$.  In \eqref{eq:forwardmodel_generic}, the matrix $\mathbf{N}$ stands for measurement noise and any mismodeling.

\begin{remark}[Complementary acquisitions]
In most cases and particularly in the applications considered in this paper (see Section \ref{sec:appli}), the number of acquisitions is limited to $K=2$ and the acquired images are of complementary spatial and spectral resolutions. One of these two images corresponds to a low spatial and high spectral resolution image and, to be more explicit, will be denoted as $\mathbf{Y}_{\mathrm{HS}}$, while the other, denoted as $\mathbf{Y}_{\mathrm{HR}}$ is of high spatial and low spectral resolution, with $B_{\mathrm{HS}} \geq B_{\mathrm{HR}}$ and $N_{\mathrm{HS}} \leq N_{\mathrm{HR}}$.
\end{remark}

This paper addresses the problem of recovering the latent image $\mathbf{X}$, which is generally of the highest spatial and spectral resolutions, i.e., $N \geq \max_k\{N_k\}$ and $B \geq\max_k\{B_k\}$. This can be formulated as the optimization problem
\begin{equation}\label{eq:optim1}
     \min_{\mathbf{X}} \left\|\mathcal{H}(\obsset) - \mathcal{M}(\mathbf{X})\right\|_{\mathrm{F}}^2 + \mathcal{R}(\mathbf{X})
\end{equation}
where $\mathcal{R}(\cdot)$ is a regularization. This penalization function is often designed to be separable with respect to the spatial and the spectral information, i.e.,
\begin{equation}\label{eq:full_regularization}
    \mathcal{R}(\mathbf{X}) = \mathcal{R}_{\mathrm{spa}}(\mathbf{X}) + \mathcal{R}_{\mathrm{spe}}(\mathbf{X})
\end{equation}
where the two terms on the right-hand side encode the expected spatial and spectral properties of $\mathbf{X}$, respectively. In most of the works dedicated to the restoration of multiband images, the pixels $\mathbf{x}_n \in \mathbb{R}^B$ ($n=1,\ldots,N$) of the unknown image $\mathbf{X}$ are assumed to live in a subspace $\mathbb{V} \subset \mathbb{R}^{\tilde{B}}$ of significantly lower dimension than the original space, i.e. $\tilde{B} \ll B$. This property can be promoted by choosing a spectral regularization $\mathcal{R}_{\mathrm{spe}}(\cdot)$ enforcing a low-rank structure on $\mathbf{X}$ by penalizing the rank
\begin{equation}
    \mathcal{R}_{\mathrm{spe}}(\mathbf{X}) = \lambda_{\mathrm{spe}}\ \mathrm{rank}\left(\mathbf{X}\right)
\end{equation}
or its convex relaxation, i.e., the nuclear norm
\begin{equation}
    \mathcal{R}_{\mathrm{spe}}(\mathbf{X}) =  \lambda_{\mathrm{spe}}\left\|\mathbf{X}\right\|_*
\end{equation}
where $\lambda_{\mathrm{spe}}$ is a hyperparameter adjusting the weight of the regularization. One data-driven alternative consists in estimating the signal subspace $\mathbb{V}$ and its dimension $\tilde{B}$ beforehand from the image of highest spectral resolution $\mathbf{G}_{\mathrm{HS}}$ available in the set $\obsset$, i.e., $\mathbf{G}_{\mathrm{HS}}\in \obsset$. This subspace estimation is generally conducted by a principal component analysis \cite{wei2015hyperspectral, dian2020regularizing} or by using a dedicated subspace identification strategy \cite{bioucas2008hyperspectral, green1988transformation}. Then the spectral regularization could be defined as
\begin{equation}
    \mathcal{R}_{\mathrm{spe}}(\mathbf{X}) = \sum_{n=1}^N \iota_{\mathbb{V}}(\mathbf{x}_n)
\end{equation}
where $\iota_{{\mathbb{V}}}(\cdot)$ is the indicator function on the set $\mathbb{V}$. However, to simultaneously reduce the computational complexity of the resulting algorithms, a widely admitted strategy consists in imposing the factorization
\begin{equation}\label{eq:regul_spectral}
    \mathbf{X}=\mathbf{V}\mathbf{A}
\end{equation}
where $\mathbf{V} \in \mathbb{R}^{B\times \tilde{B}}$ is a matrix whose $\tilde{B} \ll B$ columns span the lower dimensional subspace $\mathbb{V}$ and is generally chosen as orthonormal, i.e., $\mathbf{V}^{\top}\mathbf{V}=\mathbf{I}_{\widetilde{B}}$ where $\mathbf{I}_{\widetilde{B}}$ is the $\widetilde{B}\times \widetilde{B}$ identity matrix. The matrix $\mathbf{A} \in \mathbb{R}^{\tilde{B} \times N}$ contains the unknown representation coefficients of the pixels projected onto the subspace. Under this constraint, the original formulation \eqref{eq:optim1} can be rewritten as an optimization problem with respect to the representation coefficients $\mathbf{A}$
\begin{equation}\label{eq:optim2}
    \hat{\mathbf{A}}=\arg \min_{\mathbf{A}} \left\|\mathcal{H}(\obsset) - \mathcal{M}(\mathbf{V}\mathbf{A})\right\|_{\mathrm{F}}^2 + \mathcal{R}_{\mathrm{spa}}(\mathbf{A})
\end{equation}
with $\hat{\mathbf{X}}=\mathbf{V}\hat{\mathbf{A}}$. This latest formulation adopted by plenty of research works from the literature relies on an explicit data-driven spectral regularization specifically learnt from the observed image $\mathbf{G}_{\mathrm{HS}}$ of highest spectral resolution which acts as a spectral guidance image. Conversely, very few attempts have been dedicated to the design of a data-driven spatial regularization $\mathcal{R}_{\mathrm{spa}}(\cdot)$ exploiting the observed image of highest spatial resolution among the set of observations $\obsset$. This paper aims at filling this gap by proposing a generic framework able to encode relevant spatial information into a deep generative model acting as a regularizer. This framework is described in the next section.

\section{Proposed Framework}
\label{sec:proposed}
This section describes the general framework specifically proposed to spatially regularize multiband image inverse problems when a high spatial resolution image is available in the set $\obsset$. This image, denoted $\mathbf{G}_{\mathrm{HR}}\in \obsset$,  acts as spatial guidance image for the data-driven spatial regularization. Its generic formulation and the corresponding algorithmic scheme are introduced in Sections \ref{subsec:generic_formulation} and \ref{subsec:optimization}. This framework offers the possibility of embedding any existing deep generative model whose key feature is its ability to extract relevant spatial features from the spatial guidance image $\mathbf{G}_{\mathrm{HR}}\in \obsset$. It is worth noting that the choice of this network is left to the end-user who could select the most appropriate and up-to-date from the latest literature. In what follows, this framework is instantiated for one particular network as an illustrative purpose. Its architecture and the training strategy are detailed in Section ~\ref{sec_net}.

\subsection{Generic Formulation}\label{subsec:generic_formulation}
Inspired by the so-called deep generative model~\cite{turhan2018recent} and deep image prior approach~\cite{ulyanov2018deep}, the proposed framework leverages on the ability of deep networks to encode prior knowledge.
More precisely, a generative decoder is trained to learn a mapping $\mathsf{D}(\cdot)$ from a latent space $\mathcal{Z}$ to the space $\mathbb{R}^{\tilde{B}\times N}$ of the representation coefficients $\mathbf{A}$.  As the image $\mathbf{X}$ to be recovered is constrained to belong to the range $\mathbb{V}$ of the matrix $\mathbf{V}$ (see \eqref{eq:regul_spectral}), its representation coefficients $\mathbf{A}$  are assumed to belong to the range of the nonlinear mapping $\mathsf{D}(\cdot)$. This constraint can be satisfied by imposing
\begin{equation}
    \mathbf{A} = \mathsf{D}(\mathbf{Z})
\end{equation}
where $\mathbf{Z}\in\mathbb{R}^{k\times N}$ is the latent representation matrix equipped with a Gaussian  prior. Finally, the unknown image is estimated following
\begin{equation}
    \hat{\mathbf{X}}=\mathbf{V}\mathsf{D}( \hat{\mathbf{Z}})
\end{equation}
where the estimated latent representation matrix $\hat{\mathbf{Z}}$ is the solution of the problem
\begin{equation}\label{eq:optim2_1}
    \min_{\mathbf{Z}} \left\|\mathcal{H}(\obsset) - \mathcal{M}(\mathbf{V}\mathsf{D}(\mathbf{Z}))\right\|_{\mathrm{F}}^2 + \lambda \|\mathbf{Z}\|_{\mathrm{F}}^2
\end{equation}
with $\lambda$ a hyperparameter. The generic algorithmic scheme implemented to solve this problem is detailed in what follows.

\subsection{Optimization}\label{subsec:optimization}
The optimization problem \eqref{eq:optim2_1} can be challenging to solve, not only because of the nonlinearity induced by the mapping $\mathsf{D}(\cdot)$ but also because of the forward modeling $\mathcal{M}(\cdot)$. However it can be tackled efficiently by designing ADMM which allows the original problem to be decomposed into a 3-step procedure with simpler subproblems. By explicitly introducing the representation coefficient matrix $\mathbf{A}$, an equivalent constrained formulation writes
\begin{equation}\label{eq:optim3}
    \min_{\mathbf{A },\mathbf{Z}} \left\|\mathcal{H}(\obsset) - \mathcal{M}(\mathbf{V}\mathbf{A})\right\|_{\mathrm{F}}^2 + \lambda \|\mathbf{Z}\|_{\mathrm{F}}^2 \ \text{s.t.} \ \mathbf{A} = \mathsf{D}(\mathbf{Z}).
\end{equation}
Then the ADMM consists in iteratively performing the 3 following steps
\begin{align}
  \mathbf{A }^{(t+1)}& = \arg\min_{{\mathbf{A }}}\left\|\mathcal{H}(\obsset)-\mathcal{M}\left(\mathbf{V}\mathbf{A }\right)\right\|_{\mathrm{F}}^{2} \label{eq_lsgd_la}\\
   &\qquad \qquad \qquad +\mu\left\|\mathsf{D}(\mathbf{Z}^{(t)})-{\mathbf{A }}
  +\frac{1}{2\mu}{\mathbf{U }^{(t)}}\right\|_{\mathrm{F}}^{2} \nonumber \\
  \mathbf{Z}^{(t+1)}&=\arg\min_{\mathbf{Z}}\left\|\mathsf{D}(\mathbf{Z})-{\mathbf{A }}
  +\frac{1}{2\mu}{\mathbf{U }^{(t)}}\right\|_{\mathrm{F}}^{2}+\frac{\lambda}{\mu}\left\|\mathbf{Z}\right\|_{\mathrm{F}}^{2} \label{eq_lsgd_lz}\\
  {\mathbf{U }^{(t+1)}}& = {\mathbf{U }^{(t)}}+2\mu\left(\mathsf{D}(\mathbf{Z}^{(t+1)})-{\mathbf{A }^{(t+1)}}\right)\label{eq_lsgd_lg}
\end{align}
where ${\mathbf{U }}$ is a Lagrangian multiplier and $\mu$ is a penalty parameter. Interestingly, the minimization \eqref{eq_lsgd_la} stands for a generic formulation of an $\ell_2$-regularized inverse problem. For most multiband imaging tasks, a closed-form solution can be implemented straightforwardly, as it will be shown for two ubiquitous tasks considered in Section \ref{sec:appli}. The problem \eqref{eq_lsgd_lz} is a nonlinear least-square problem similar to a projection onto the range of $\mathsf{D}(\cdot)$. In practice, it is empirically solved by resorting to a optimizer dedicated to deep learning, e.g., Adam. The overall algorithmic sketch of the proposed generic framework is summarized in Algorithm~\ref{alg:framework}.

\begin{algorithm}
	\renewcommand{\algorithmicrequire}{\textbf{Input:}}
	\renewcommand{\algorithmicensure}{\textbf{Output:}}
	\caption{Multiband image inversion: generic formulation}
	\label{alg:framework}
	\begin{algorithmic}[1]
	\REQUIRE Set $\obsset$ of observed images, regularization parameters $\lambda$ and $\mu$.
    \STATE Identify a basis $\mathbf{V}$ of the spectral subspace using PCA.
    \STATE Train the deep generative decoder $\mathsf{D}(\cdot)$.
	\STATE Initialization: ${\mathbf{A }}$, $\mathbf{Z}$ and ${\mathbf{U }}$ with zeros.
    \STATE \textbf{while} not converged \textbf{do}:
    \STATE \hspace{2mm} Update ${\mathbf{A }}$ using~\eqref{eq_lsgd_la};
    \STATE \hspace{2mm} Update $\mathbf{Z}$ using~\eqref{eq_lsgd_lz};
    \STATE \hspace{2mm} Update ${\mathbf{U }}$ using~\eqref{eq_lsgd_lg};
    \STATE \textbf{end while}
    \STATE $\hat{\mathbf{X}}=\mathbf{V}\mathsf{D}(\mathbf{Z})$.
    \ENSURE Estimated multiband image $\hat{\mathbf{X}}$.
  \end{algorithmic}
\end{algorithm}

\subsection{ Guided deep decoder based generative model}
\label{sec_net}
As an illustrative instance, one particular network from the literature is considered to learn the generative model $\mathsf{D}(\cdot)$ from the spatial guidance image $\mathbf{G}_{\mathrm{HR}}$. This guided deep decoder proposed in~\cite{uezato2020guided} is designed to span the space of the representation coefficients $\mathbf{A}$ from a low-dimensional manifold latent variables $\mathbf{Z}$ with the spatial prior information  encoded by their parameters.

\begin{figure}
  \centering
  \includegraphics[width=8.5cm]{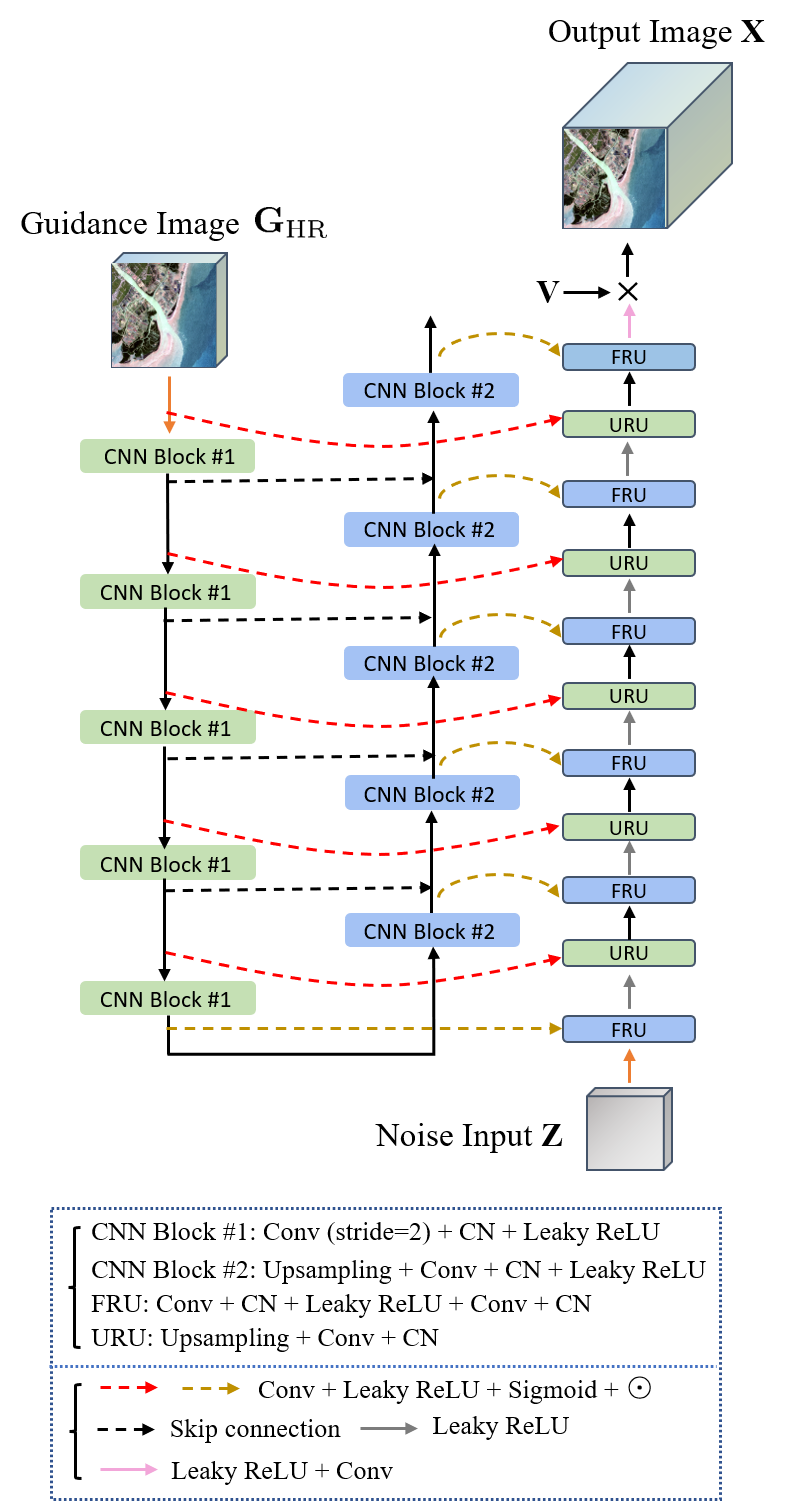}\\
  \caption{The architecture of the guided deep generative model.}\label{fig_frame_network_GDD}
\end{figure}

The network consists of two streams. The first one is a U-net based encoder-decoder architecture while the second one is a deep decoder and comprises upsampling refinement units (URU) and feature refinement units (FRU). The whole architecture is depicted in Fig.~\ref{fig_frame_network_GDD}. Inspired by the design of deep image prior architectures~\cite{ulyanov2018deep}, the deep decoder is trained to map a random generated noise $\mathbf{Z}$ to the estimated subspace coefficients $\mathbf{A}=\mathsf{D}(\mathbf{Z})$. The image structure information is encoded in the network parameters which can be resorted as an implicit image prior. The particularity of the GDD is the following: the input of the encoder-decoder network is assumed to be the auxiliary image of highest spatial resolution. It plays the role of guidance image whose spatial features are extracted at different scale to be used as conditional weights to guide the deep decoder.  This model is trained by minimizing an energy function, here chosen as the mean square error
\begin{equation}\label{eq.loss_GDD}
  \mathcal{L}_{\text{GDD}} = \|\mathcal{H}(\obsset) -\mathcal{M}\left(\mathbf{VA}\right)\|_\text{F}^{2}.
\end{equation}
Once trained, the decoder can be used as the generator $\mathsf{D}(\cdot)$.


\section{Applications}
\label{sec:appli}
To demonstrate the versatility of the proposed method to tackle challenging multiband imaging inverse problems, it is instantiated for two ubiquitous tasks, namely image fusion and multiband image inpainting. The instances associated to these two applications are detailed in what follows.

\subsection{Multiband image fusion}\label{subsec:app_fusion}

\subsubsection{Problem formulation}
Given a pair of observed images $\obsset= \left\{\mathbf{Y}_{\mathrm{HR}}, \mathbf{Y}_{\mathrm{HS}}\right\}$ of complementary resolutions, multiband image fusion  aims to fuse a high spatial and low spectral resolution image $\mathbf{Y}_{\mathrm{HR}}$ with a low spatial resolution and high spectral resolution  image $\mathbf{Y}_{\mathrm{HS}}$. The fused image $\mathbf{X}$ is expected to be of the highest spatial and spectral resolutions, $B=B_{\mathrm{HS}}$ and $N=N_{\mathrm{HR}}$. When the high spatial resolution image is a panchromatic image $\mathbf{Y}_{\mathrm{HR}}$ (i.e., $B_{\mathrm{HR}}=1$), typical scenarios of this problem are referred to as pansharpening or hyperspectral pansharpening if the complementary image is a multispectral or hyperspectral images, respectively. The generic model \eqref{eq:forwardmodel_generic} can then be instantiated by specifying the operators $\mathcal{H}(\cdot)$ and $\mathcal{M}(\cdot)$ as
\begin{equation}
    \mathcal{H}(\obsset) = \begin{bmatrix}\mathbf{Y}_{\mathrm{HS}} \\ \mathbf{Y}_{\mathrm{HR}} \end{bmatrix} \ \text{and} \
     \mathcal{M}(\mathbf{X}) = \begin{bmatrix}\mathbf{X}\mathbf{BS} \\ \mathbf{R}\mathbf{X} \end{bmatrix}
\end{equation}
where  $\mathbf{B}\in\mathds{R}^{N\times N}$ is a cyclic convolution operator which stands for a spatial blurring, $\mathbf{S}$ denotes a regular spatial downsampling matrix with a downsampling factor denoted by $f$ and $\mathbf{R}\in\mathbb{R}^{\tilde{B}\times B}$ is a spectral response. In most works of the literature \cite{wei2015hyperspectral,wei2015fast}, the spectral regularization \eqref{eq:regul_spectral} is designed after conducting a principal component analysis of the high spectral resolution image, i.e. the spectral guidance image is chosen as $\mathbf{G}_{\mathrm{HS}}=\mathbf{Y}_{\mathrm{HS}}$. Conversely, the spatial regularization \eqref{eq:full_regularization} is generally chosen as a parametric model promoting expected spatial characteristics, e.g., total variation \cite{simoes2014convex} or Sobolev  \cite{guilloteau2020hyperspectral} for promoting piece-wise constant or smooth patterns, respectively. Instead, we propose to explicitly spatially regularize the fusion problem by resorting to the high spatial resolution image as a spatial guidance image, i.e., $\mathbf{G}_{\mathrm{HR}}=\mathbf{Y}_{\mathrm{HR}}$. The optimization problem \eqref{eq:optim2_1} can be rewritten as
\begin{equation}\label{eq_sr_lsgd}
   \min_{\mathbf{Z}}\|\mathbf{Y}_{\textrm{HS}}-\mathbf{V}\mathsf{D}(\mathbf{Z})\mathbf{B}\mathbf{S}\|_\text{F}^{2}
   + \|\mathbf{Y}_{\textrm{HR}}-\mathbf{R}\mathbf{V}\mathsf{D}(\mathbf{Z})\|_\text{F}^{2}
   +\lambda\|\mathbf{Z}\|_\text{F}^{2}.
\end{equation}
The generic algorithm proposed in Section \ref{subsec:optimization} to solve \eqref{eq_sr_lsgd} is instantiated below.

\subsubsection{Optimization}
Introducing ${\mathbf{A }}=\mathsf{D}(\mathbf{Z})$, the augmented Lagrangian function can be written as
\begin{equation}\label{eq_sr_lsgd_l}
\begin{aligned}
\mathcal{L}\left({{\mathbf{A }},\mathbf{Z},{\mathbf{U }}}\right)
  &=\left\|\mathbf{Y}_{\textrm{HS}}-\mathbf{V}{\mathbf{A }}\mathbf{B}\mathbf{S}\right\|_\text{F}^{2}
   + \left\|\mathbf{Y}_{\textrm{HR}}-\mathbf{R}\mathbf{V}{\mathbf{A }}\right\|_\text{F}^{2}\\
  &+\mu\left\|\mathsf{D}(\mathbf{Z})-{\mathbf{A }}
  +\frac{{\mathbf{U }}}{2\mu}\right\|_\text{F}^{2}+\lambda\left\|\mathbf{Z}\right\|_\text{F}^{2}.
  \end{aligned}
\end{equation}
Updating ${\mathbf{A}}$ according to the rule \eqref{eq_lsgd_la} is a strongly convex problem that can be solved analytically by forcing the corresponding gradient to be zero. An efficient implementation of the solution can be derived following the strategy proposed by \cite{wei2015fast}. By noting that $\mathbf{V}$ is an orthogonal matrix with $\mathbf{V}^{\top}\mathbf{V}=\mathbf{I}_{\widetilde{B}}$, it consists in solving
the Sylvester equation
\begin{equation}\label{eq_sr_syl}
  \mathbf{C}_1{\mathbf{A }}+{\mathbf{A }}\mathbf{C}_2=\mathbf{C}_3.
\end{equation}
with
\begin{equation}
\begin{aligned}\label{sr_lsgd_la_syl}
&\mathbf{C}_{1}=(\mathbf{R V})^{\top}(\mathbf{R V})+\mu \mathbf{I}_{\widetilde{B}} \\
&\mathbf{C}_{2}=(\mathbf{B S})(\mathbf{B S})^{\top} \\
&\mathbf{C}_{3}=\mathbf{V}^{\top} \mathbf{Y}_{\mathrm{HS}}(\mathbf{B S})^{\top}+(\mathbf{R V})^{\top} \mathbf{Y}_{\mathrm{HR}}+\mu\left(\mathsf{D}(\mathbf{Z})+\frac{{\mathbf{U }}}{2 \mu}\right).
\end{aligned}
\end{equation}
The resulting algorithmic scheme is recalled in Algo. \ref{alg:fusion} for completeness. Regarding the updating rules for $\mathbf{Z}$ and ${\mathbf{U }}$, they can follow the derivations in~\eqref{eq_lsgd_lz} and \eqref{eq_lsgd_lg}.

\begin{algorithm}[!t]
	\renewcommand{\algorithmicrequire}{\textbf{Input:}}
	\renewcommand{\algorithmicensure}{\textbf{Output:}}
	\caption{Solution for \eqref{eq_sr_syl} by solving the Sylvester equation.}
	\label{alg:fusion}
	\begin{algorithmic}[1]
	\REQUIRE $\mathbf{Y}_{\textrm{HR}}$, $\mathbf{Y}_{\textrm{HS}}$, $\mathbf{V}$,
$\mathbf{B}$, $\mathbf{S}$, $f$, $\mathbf{R}$, $\mathsf{D}(\mathbf{Z})$, ${\mathbf{U }}$ and $\mu$.
    \STATE Compute $\mathbf{C}_1$, $\mathbf{C}_2$ and $\mathbf{C}_3$ using \eqref{sr_lsgd_la_syl}.
    \STATE Eigen-decomposition of $\mathbf{B}$: $\mathbf{B}=\mathbf{FD}\mathbf{F}^\text{H}$.
    \STATE $\overline{\mathbf{D}}=\mathbf{D}(\mathbf{1}_f\otimes \mathbf{I}_{n})$.
    \STATE Eigen-decomposition of $\mathbf{C}_1$: $\mathbf{C}_1=\mathbf{Q}\mathbf{\Lambda}\mathbf{Q}^{-1}$.
    \STATE $\overline{\mathbf{C}}_3=\mathbf{Q}^{-1}\mathbf{C}_3\mathbf{F}$.
    \STATE \textbf{for} $i=1~\text{to}~\widetilde{B}$ \textbf{do}:
    \STATE \hspace{2mm} $\overline{{\mathbf{A }}}_{i}=\lambda_{i}^{-1}\left(\overline{\mathbf{C}}_{3}\right)_{i}-\lambda_{i}^{-1}\left(\overline{\mathbf{C}}_{3}\right)_{i} \overline{\mathbf{D}}\left(\lambda_{i} f \mathbf{I}_{n}+\sum_{t=1}^{f} \mathbf{D}_{t}^{2}\right) \overline{\mathbf{D}}^{\text{H}}$
    \STATE \textbf{end for}
    \STATE $\widehat{{\mathbf{A }}}=\mathbf{Q}{\mathbf{A }}\mathbf{F}^{\text{H}}$.
    \ENSURE The estimate $\widehat{{\mathbf{A }}}$.
  \end{algorithmic}
\end{algorithm}

\subsection{Multiband image inpainting}\label{subsec:app_inpainting}
\subsubsection{Problem formulation}
Because of sensor malfunctions or miscalibrations, multiband images are often affected by so-called \emph{dead pixels}, i.e., pixels with unreliable measurements. Moreover, acquiring a multiband image for every spatial position is generally time-consuming or may damage the observed scene in particular in microscopy \cite{monier2018reconstruction}. In such cases, the full hyperspectral image $\mathbf{X}$ should be restored from a partial spatial sampling acquisition $\boldsymbol{\Omega}_{\mathrm{b}}\mathbf{Y}_{\mathrm{HS}}\boldsymbol{\Omega}_{\mathrm{p}}$ where $\boldsymbol{\Omega}_{\mathrm{b}} \in \left\{0,1\right\}^{\tilde{B} \times {B}}$ and $\boldsymbol{\Omega}_{\mathrm{p}} \in \left\{0,1\right\}^{N \times \tilde{N}}$ stand for binary matrices acting as masks to identify the ${\tilde{B}}$ out of $B$ non-corrupted bands and the ${\tilde{N}}$ out of $N$  non-corrupted pixels ${\tilde{N}} \leq N$. This task is referred to as hyperspectral inpainting. In some applicative scenarios, this multiband image sensing can be easily complemented by a auxiliary acquisition of a full  image $\mathbf{Y}_{\mathrm{HR}}$ at lower spectral resolution. For instance, in the context of hyperspectral imaging, this complementary image can be a RGB image ($\tilde{B}=3$) \cite{yan2020reconstruction}. When dealing with EELS spectro-microscopy, the acquisition of the EELS data can be easily preceded by annular dark-field (HAADF) imaging providing a single-band image ($\tilde{B}=1$) with a full spatial resolution. The set of available images  is then $\obsset = \left\{\mathbf{Y}_{\mathrm{HS}}, \mathbf{Y}_{\mathrm{HR}}\right\}$. The multiband image $\mathbf{Y}_{\mathrm{HS}}$ and  the auxiliary image $\mathbf{Y}_{\mathrm{HR}}$ serve here as the spectral and spatial guidance images, respectively, i.e.,  $\mathbf{G}_{\mathrm{HS}}=\mathbf{Y}_{\mathrm{HS}}$ and $\mathbf{G}_{\mathrm{HR}}=\mathbf{Y}_{\mathrm{HR}}$. Then
the generic model \eqref{eq:forwardmodel_generic} can be instantiated by defining the operators $\mathcal{H}(\cdot)$ and $\mathcal{M}(\cdot)$ as
\begin{equation}
    \mathcal{H}(\obsset) = \boldsymbol{\Omega}_\mathrm{b}\mathbf{Y}_{\mathrm{HS}}\boldsymbol{\Omega}_\mathrm{p} \ \text{and} \
     \mathcal{M}(\mathbf{X}) = \boldsymbol{\Omega}_\mathrm{b}\mathbf{X}\boldsymbol{\Omega}_\mathrm{p}.
\end{equation}
The generic problem \eqref{eq:optim2_1} can be rewritten as
\begin{equation}\label{eq_inp_lsgd}
   \min_{\mathbf{Z}}\|\boldsymbol{\Omega}_\mathrm{b}\mathbf{Y}_{\mathrm{HS}}\boldsymbol{\Omega}_\mathrm{p}-{\Omega}_\mathrm{b}\mathbf{V}\mathsf{D}(\mathbf{Z})\boldsymbol{\Omega}_\mathrm{p}\|_\text{F}^{2}+\lambda\|\mathbf{Z}\|_\text{F}^{2}.
\end{equation}
The next section details the corresponding instance of the generic algorithm proposed in Section \ref{subsec:optimization} to solve \eqref{eq_sr_lsgd}.

\subsubsection{Optimization}
The augmented Lagrangian associated with \eqref{eq_inp_lsgd} writes
\begin{equation}\label{eq_inp_lsgd_l}
\begin{aligned}
\mathcal{L}\left({{\mathbf{A }},\mathbf{Z},{\mathbf{U }}}\right)
  &=
  \left\|\boldsymbol{\Omega}_\mathrm{b}\mathbf{Y}_{\mathrm{HS}}\mathbf{\Omega}_\mathrm{b}-\boldsymbol{\Omega}_\mathrm{b}\mathbf{V}{\mathbf{A }}\mathbf{\Omega}_\mathrm{b}\right\|_\text{F}^{2}\\
  &+\mu\left\|\mathsf{D}(\mathbf{Z})-{\mathbf{A }}
  +\frac{{\mathbf{U }}}{2\mu}\right\|_\text{F}^{2}+\lambda\left\|\mathbf{Z}\right\|_\text{F}^{2},
  \end{aligned}
\end{equation}
where ${\mathbf{A }}=\mathsf{D}(\mathbf{Z})$ denotes the auxiliary variable, ${\mathbf{U }}$ is the Lagrangian multiplier. Updating the auxiliary variable ${\mathbf{A }}$ boils down to solving the quadratic problem
\begin{equation}\label{eq_inp_lsgd_v}
 \min_{{\mathbf{A }}}\left\|\boldsymbol{\Omega}_\mathrm{b}\mathbf{Y}_{\mathrm{HS}}\mathbf{\Omega}_\mathrm{p}-\boldsymbol{\Omega}_\mathrm{b}\mathbf{V}{\mathbf{A }}\mathbf{\Omega}_\mathrm{p}\right\|_\text{F}^{2}+\mu\left\|\mathsf{D}(\mathbf{Z})-{\mathbf{A }}
  +\frac{{\mathbf{U }}}{2\mu}\right\|_\text{F}^{2}.
\end{equation}
Because of the large size of this problem, its resolution is not straightforward. Inspired by the fast implementations proposed in \cite{zhuang2018fast} and \cite{lin2021admm}, the problem \eqref{eq_inp_lsgd_v} can be rewritten by vectorizing all quantities. More precisely, by denoting $\mathbf{y}_{\mathrm{HS}}=\text{vec}\{\mathbf{Y}_{\mathrm{HS}}\}$, ${\mathbf{a }}=\text{vec}\{\mathbf{A }\}$, $\mathsf{d}(\mathbf{Z})=\text{vec}\{\mathsf{D}(\mathbf{Z})\}$ and ${\mathbf{u }}=\text{vec}\{\mathbf{U }\}$ where $\text{vec}\{\cdot\}$ stacks the columns of the corresponding matrix, the problem is equivalent to
\begin{equation}\label{eq_inp_lsgd_v_2}
  \min_{{\mathbf{a }}}\left\|\mathbf{M}\mathbf{y}_{\mathrm{HS}}-\mathbf{M}\left(\mathbf{I}_{B}\otimes\mathbf{V}\right){\mathbf{a }}\right\|_\text{F}^{2}
  +\mu\left\|\mathsf{d}(\mathbf{Z})-{\mathbf{a }}
  +\frac{{\mathbf{u }}}{2\mu}\right\|_\text{F}^{2}
\end{equation}
where $\mathbf{M}\in\mathbb{R}^{\tilde{B}N \times BN}$ is the vectorization-based counterpart binary matrix of the masks $\boldsymbol{\Omega}_\mathrm{b}$ and $\boldsymbol{\Omega}_\mathrm{p}$. It yields a closed-form solution of \eqref{eq_inp_lsgd_v} given by
\begin{equation}\label{eq_inp_lsgd_v_3}
  {\mathbf{A }}= \text{vec}^{-1} \left\{\left[\mathbf{Q}\mathbf{Q}^{\top}+\mu\mathbf{I}_{BN}\right]^{-1}\left[\mathbf{Q}\mathbf{M}\mathbf{y}_{\mathrm{HS}}+\mu\mathbf{g}\right]\right\}
\end{equation}
where $\mathbf{Q}=\left(\mathbf{I}_{B}\otimes\mathbf{V}^{\top}\right)\mathbf{M}^{\top}$
and $\mathbf{g}=\mathsf{d}(\mathbf{Z})+{{\mathbf{u }}}/{2\mu}$.
The updates of $\mathbf{Z}$ and ${\mathbf{U }}$ are the same as in \eqref{eq_lsgd_lz} and \eqref{eq_lsgd_lg}.

\section{Experiments}
\label{sec:experiment}
This section shows how the proposed framework performs when tackling multiband imaging problems detailed in Section \ref{sec:appli}, namely fusion and inpainting. For each task, the performance of the proposed method, referred to as ADMM-GDD, is compared to the performances reached by dedicated state-of-the-art methods. These compared methods will be detailed in the respective sections (see Sections \ref{subsec:results_fusion} and \ref{subsec:results_inpainting}, respectively). In addition, the proposed framework is instantiated when the generative model $\mathsf{D}(\cdot)$ is not spatially informed by the guidance image $\mathbf{G}_{\mathrm{HS}}$. To do so, the guided deep decoder detailed in Section \ref{sec_net} is replaced by a variational autoencoder (VAE) trained on a generic data set. Considering this framework, referred to as ADMM-VAE, will allow to highlight the benefits of informing the generative model with the guidance image of high spatial resolution. Details regarding the architecture and the training of the VAE-based generative model are given in the Appendix. Finally, to evaluate the relevance of the splitting-based algorithm detailed in Section \ref{subsec:optimization}, Adam is used to directly solve \eqref{eq:optim2_1} instead of implementing the ADMM. The corresponding methods are coined as Adam-GDD and Adam-VAE.

\subsection{Quality Metrics}
Five figures-of-merit are used to quantitatively compare the results provided by the algorithms.
\begin{itemize}
  \item \emph{PSNR}: The peak signal-to-noise ratio (PSNR) is used to quantitatively evaluate the global similarity between the ground-truth and the estimate. It consists in computing the average single-band SNR over the bands. The bigger the better estimation.
  \item \emph{SAM}: The spectral angle mapper (SAM)~\cite{yuhas1992discrimination} is a spectral distortion metric. It is computed by averaging the SAM over the pixels. The smaller the better estimation.
  \item \emph{UIQI}: The universal image quality index (UIQI)~\cite{wang2002universal} evaluates the similarity of correlation, luminance and contrast. The single-band UIQI are averaged over the bands. The bigger the better estimation.
  \item \emph{ERGAS}: The relative dimensionless global error in synthesis (ERGAS)~\cite{wald2000quality} is a band-wise mean-normalized root-mean-square error (RMSE) which is expected to be robust to calibration. The overall ERGAS is averaged over the bands. The smaller the better estimation.
  \item \emph{SSIM}: The structural similarity index (SSIM)~\cite{wang2004image} is widely used to measure the structural similarities of the gray image. It is extended to multiband image by averaging the bandwise SSIM over all bands. The bigger the better estimation.
\end{itemize}

\begin{figure}[h]
  \centering
  \includegraphics[width=9cm]{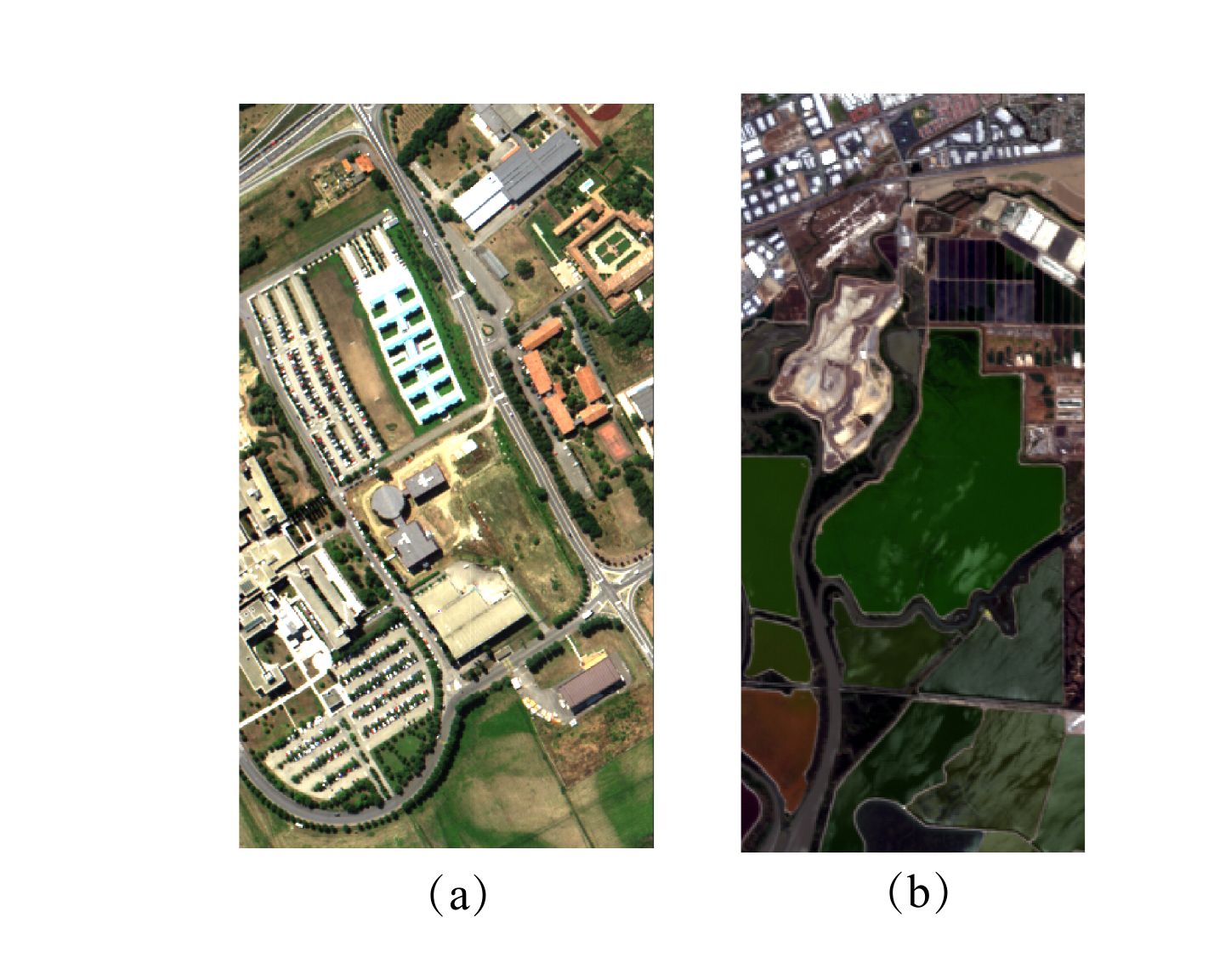}\\
  \caption{Fusion experiment -- Reference images: (a) Pavia University and (b)  Moffett field.}\label{fig_sr_RGB}
\end{figure}

\newcommand{\subfigwidth}{1.75cm}
\setlength{\tabcolsep}{1pt}
\begin{figure*}
  \centering
  \scriptsize
  \resizebox{\textwidth}{!}{
  \begin{tabular}{cccccccccc}
  HySure & FUSE-S & CMS & Deep-HS-prior & CNN-Fus & GDD & Adam-VAE & ADMM-VAE & Adam-GDD & ADMM-GDD\\
   \includegraphics[width=\subfigwidth]{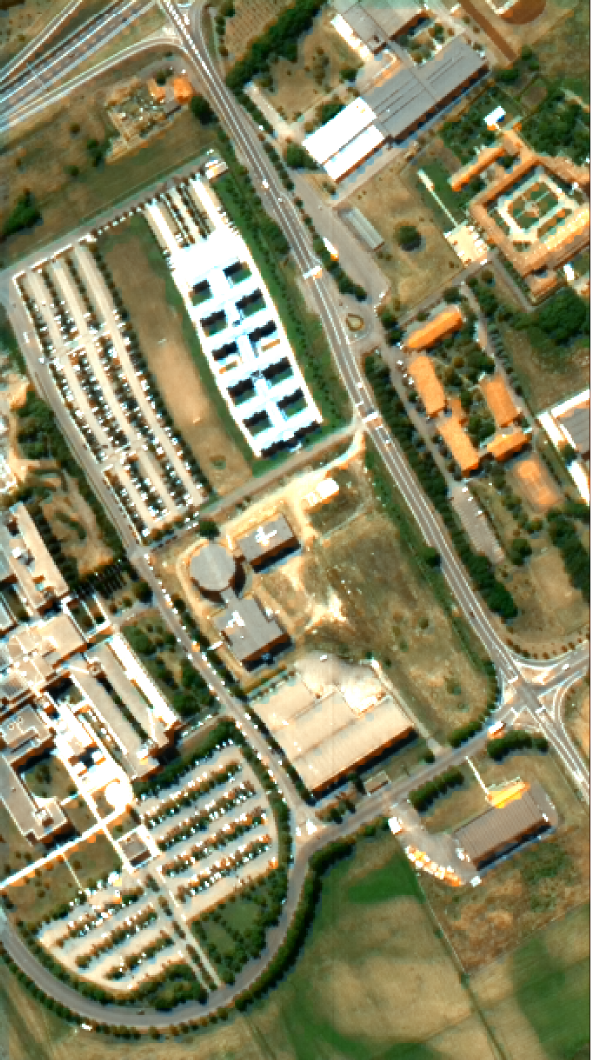} &
   \includegraphics[width=\subfigwidth]{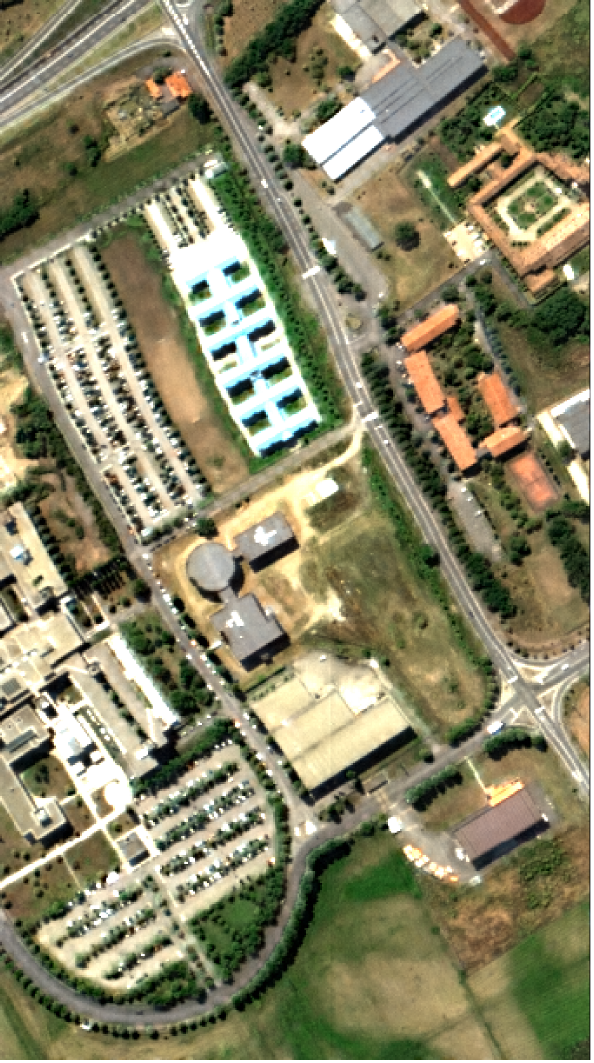} &
   \includegraphics[width=\subfigwidth]{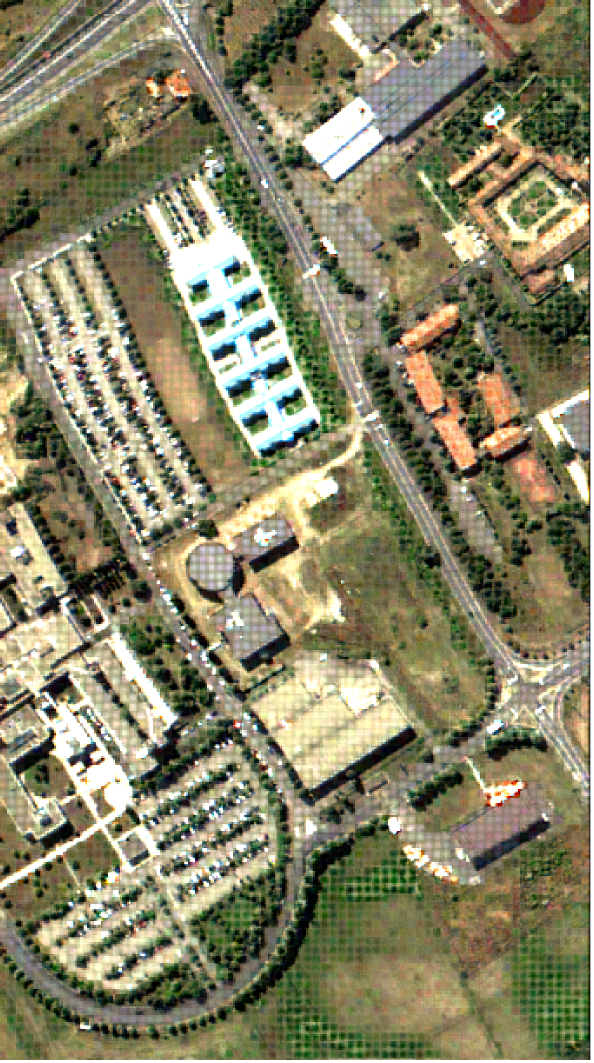} &
   \includegraphics[width=\subfigwidth]{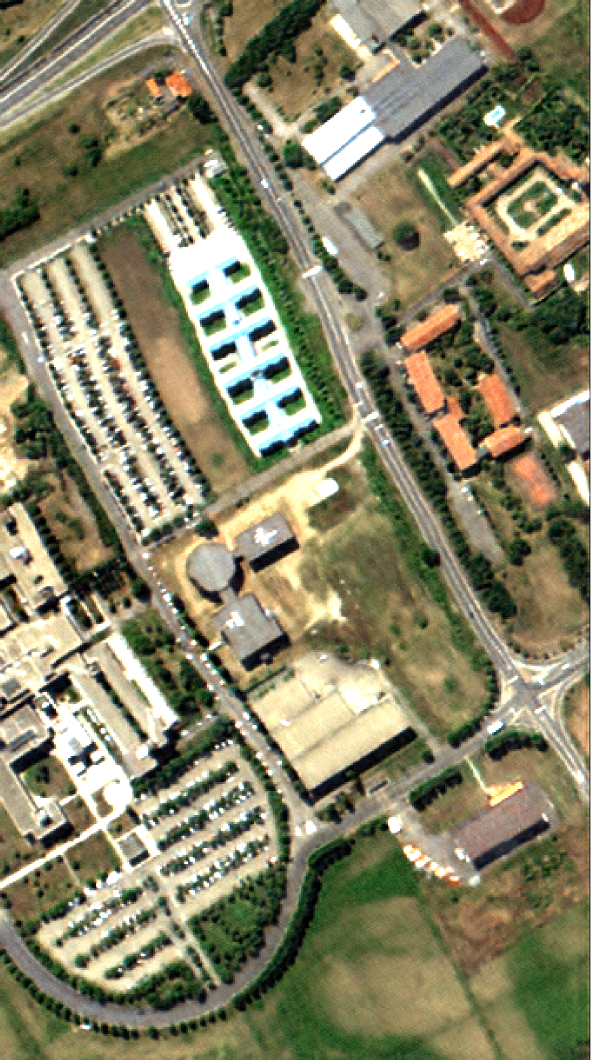} &
   \includegraphics[width=\subfigwidth]{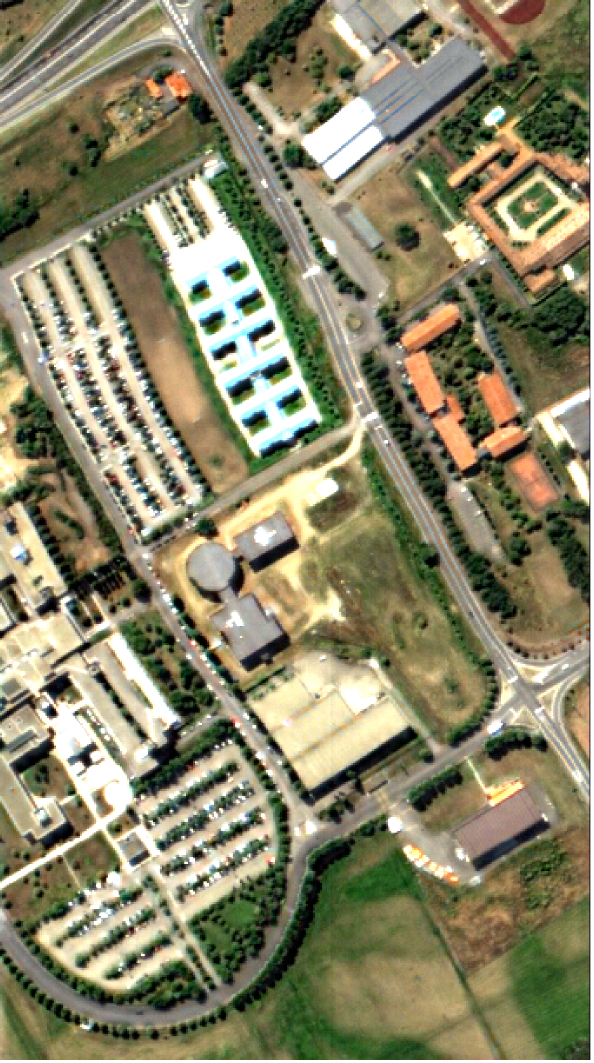} &
   \includegraphics[width=\subfigwidth]{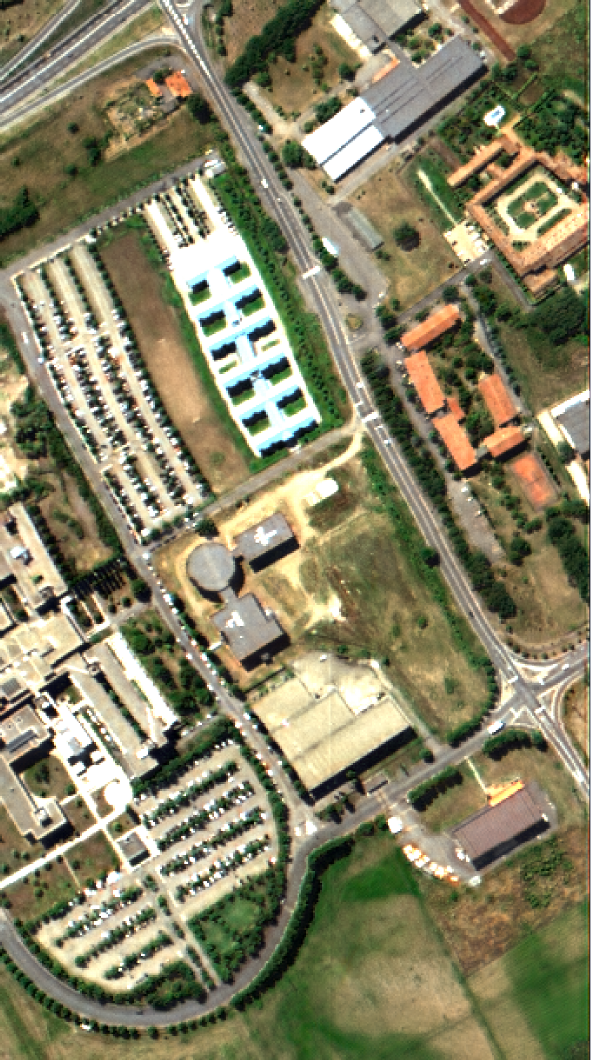} &
   \includegraphics[width=\subfigwidth]{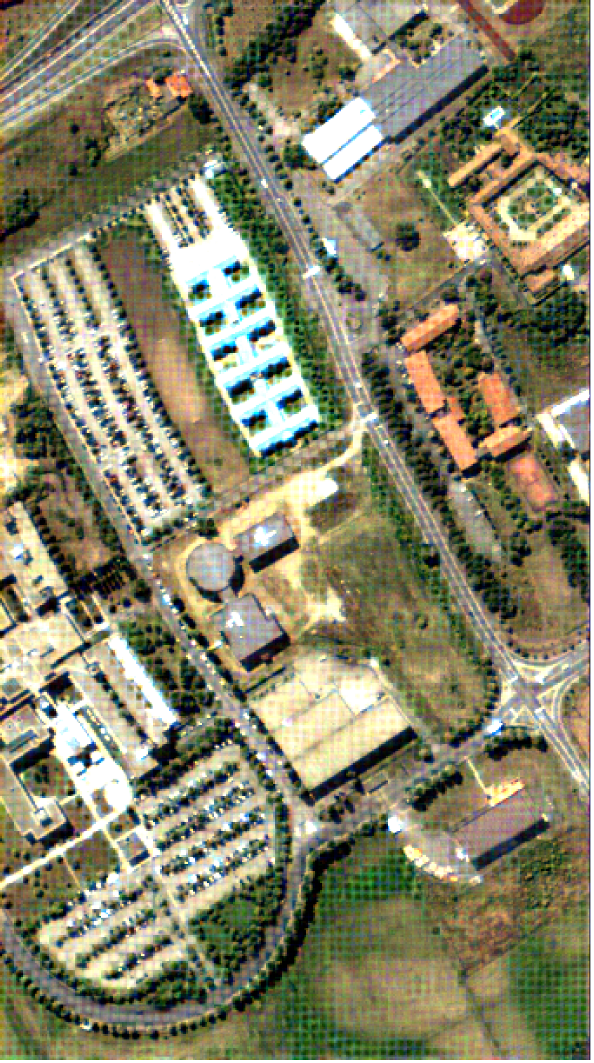} &
   \includegraphics[width=\subfigwidth]{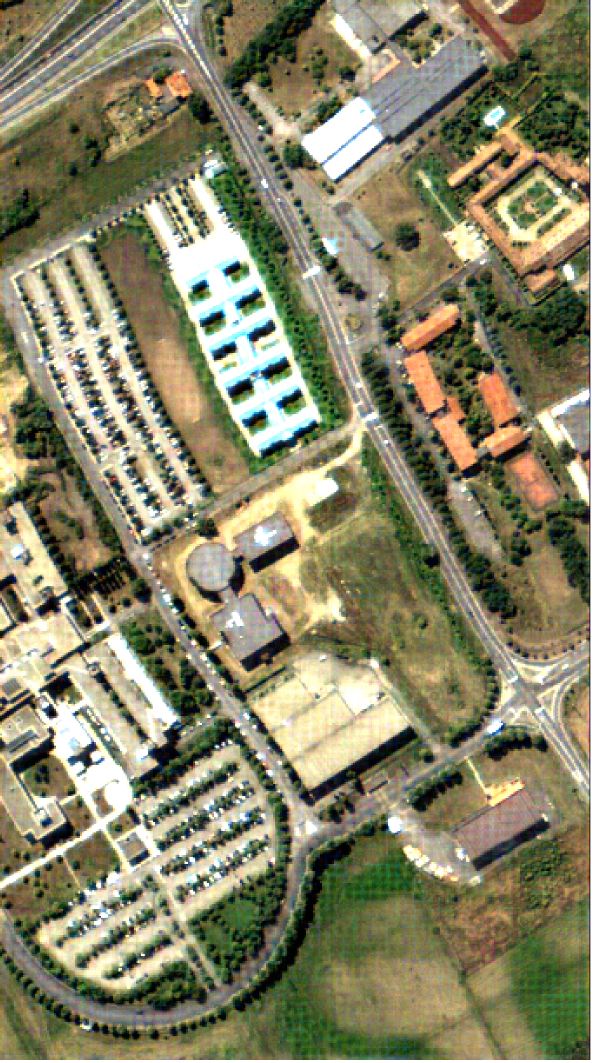} &
   \includegraphics[width=\subfigwidth]{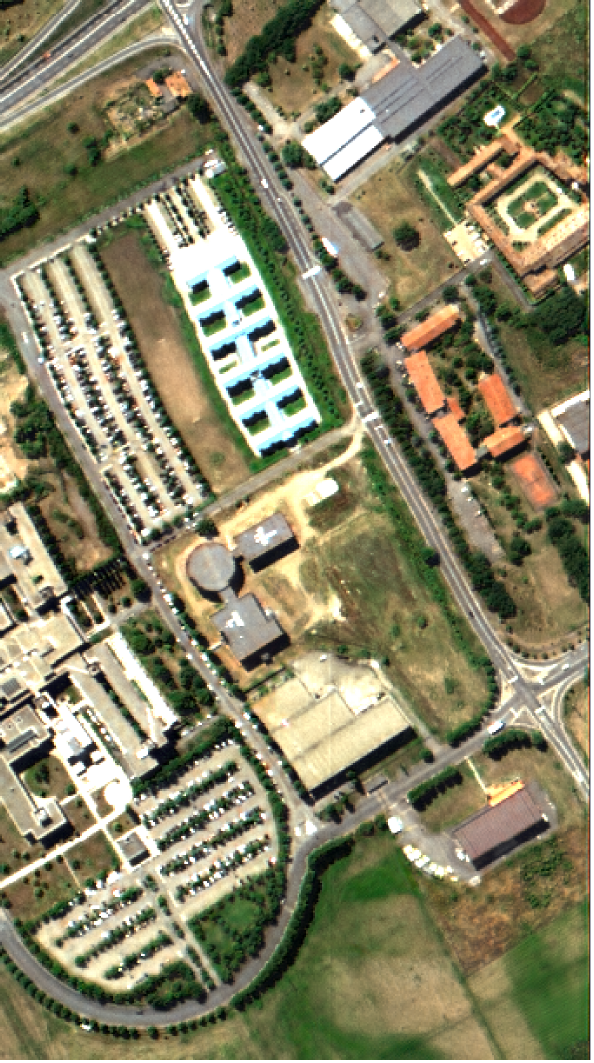} &
   \includegraphics[width=\subfigwidth]{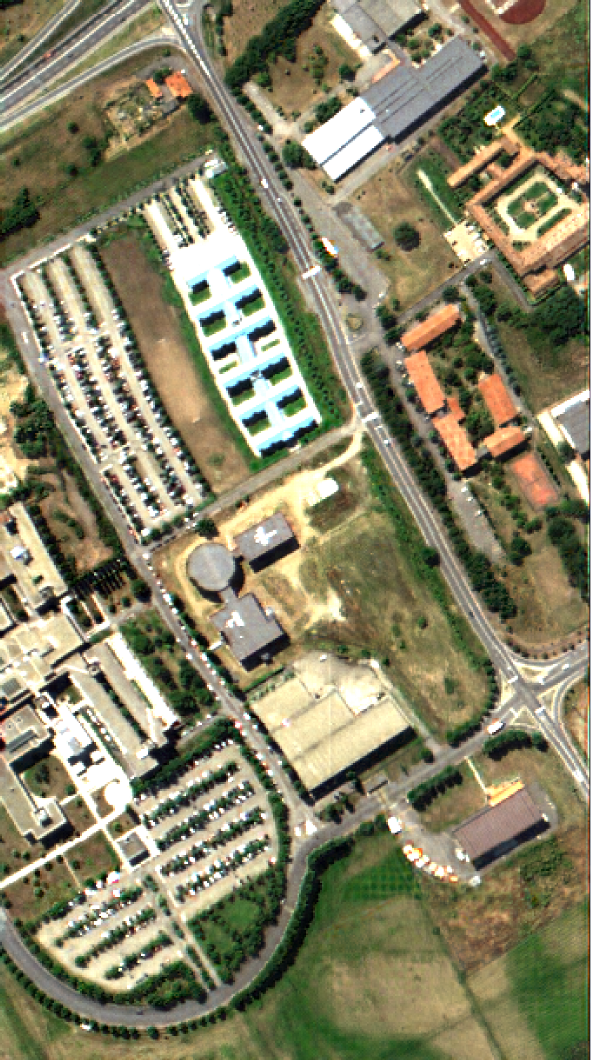}
   \\
   \includegraphics[width=\subfigwidth]{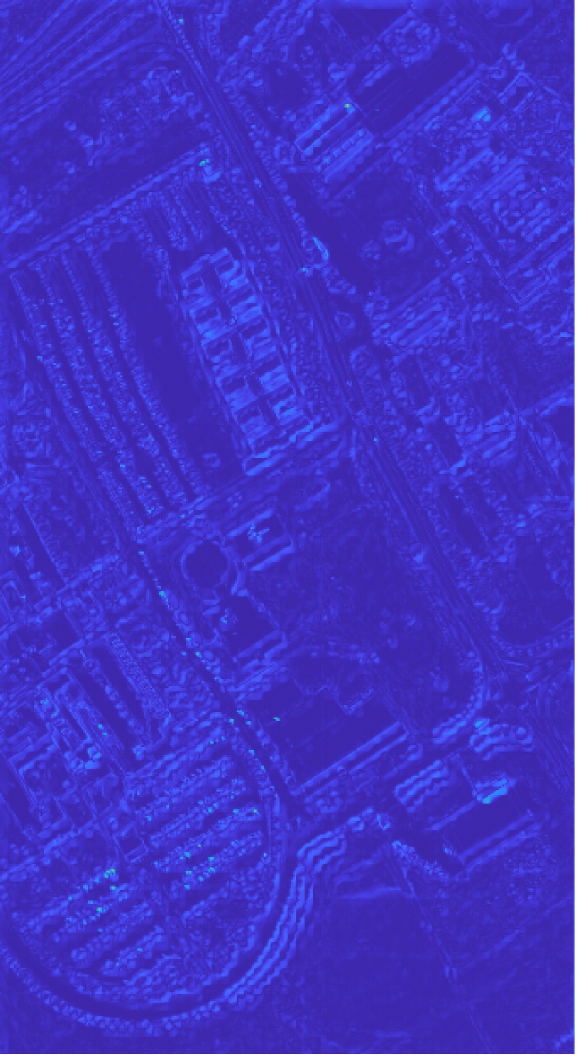} &
   \includegraphics[width=\subfigwidth]{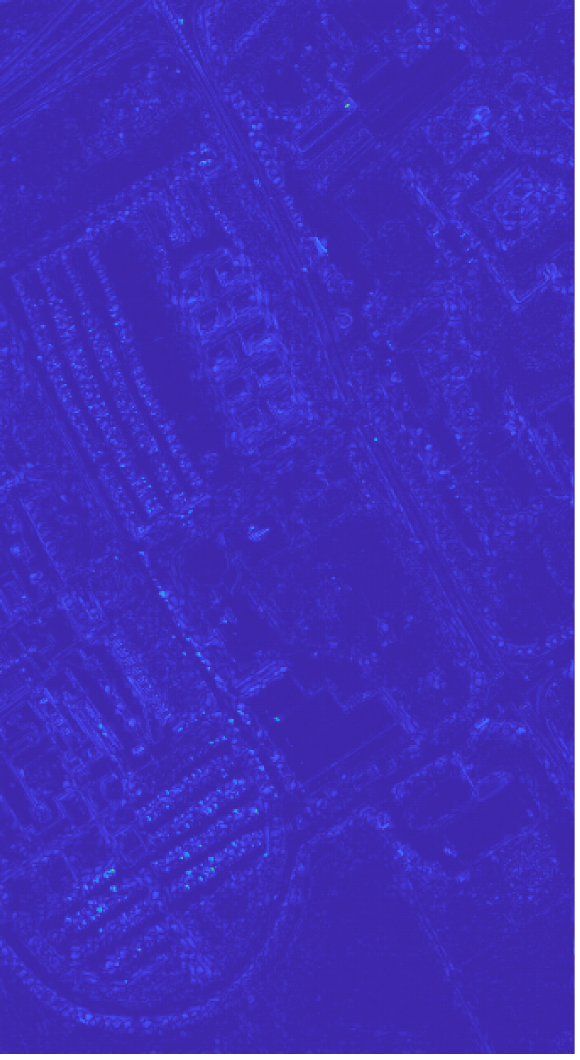} &
   \includegraphics[width=\subfigwidth]{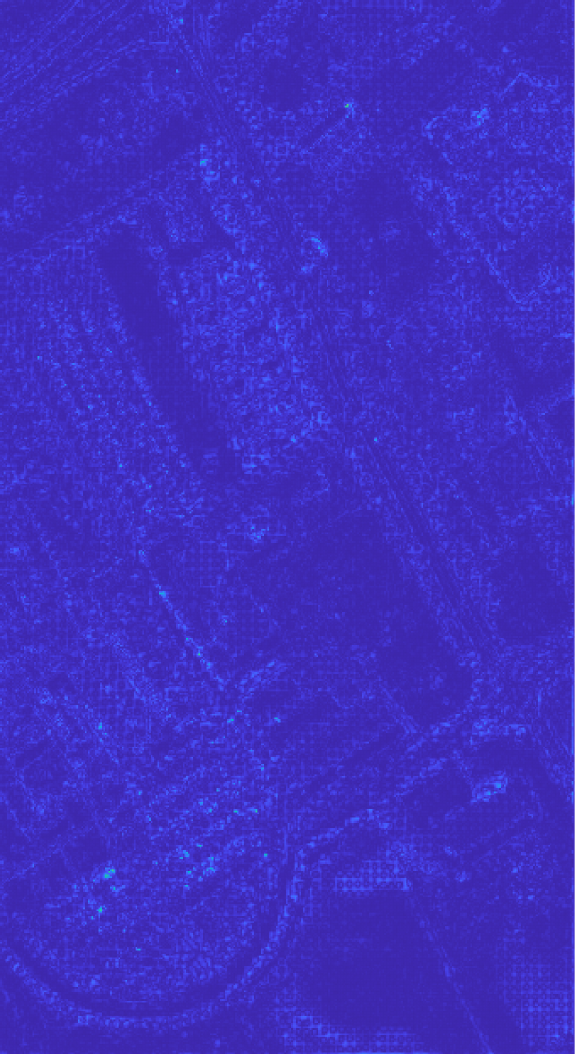} &
   \includegraphics[width=\subfigwidth]{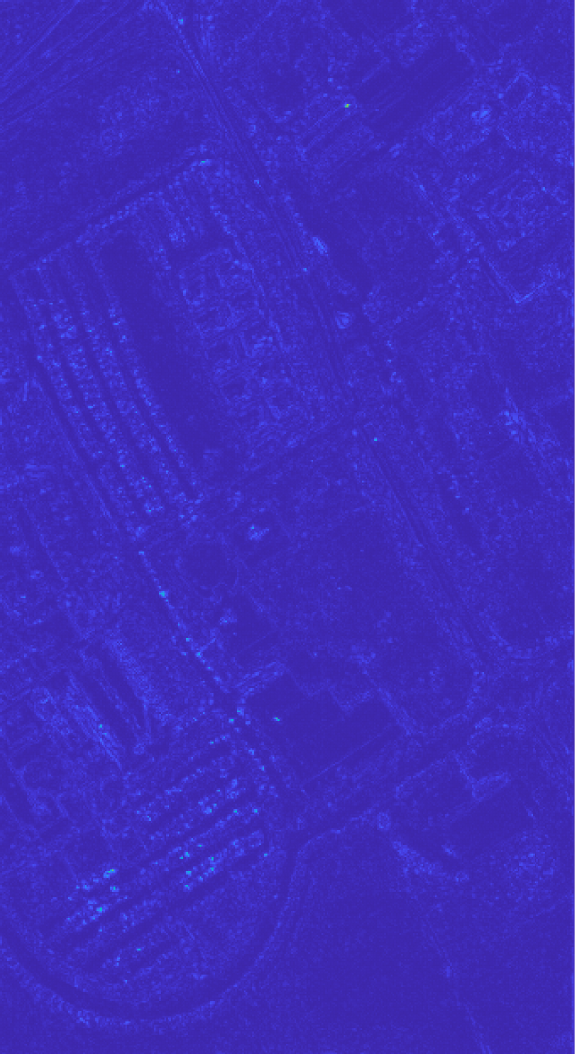} &
   \includegraphics[width=\subfigwidth]{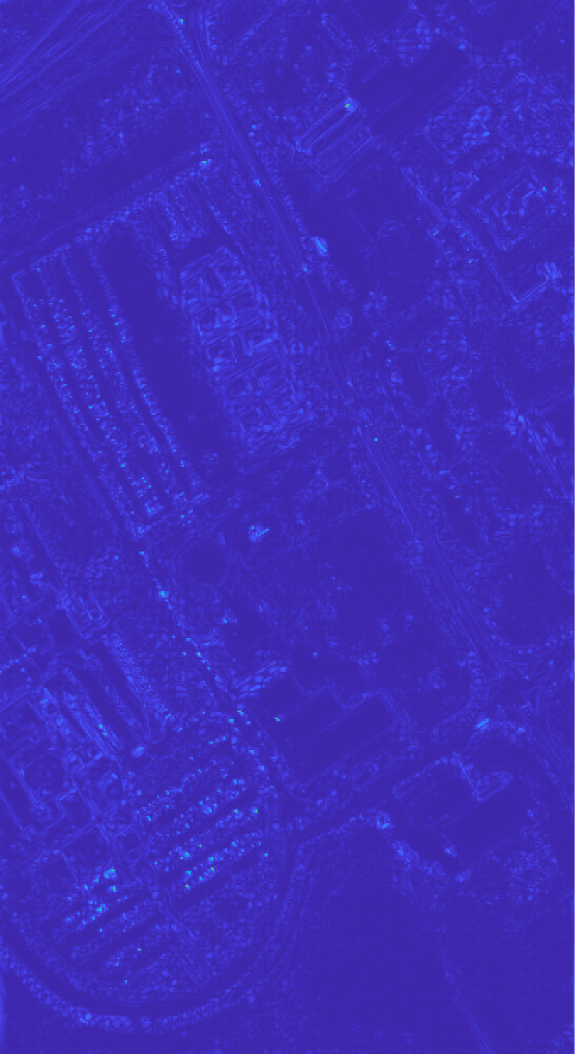} &
   \includegraphics[width=\subfigwidth]{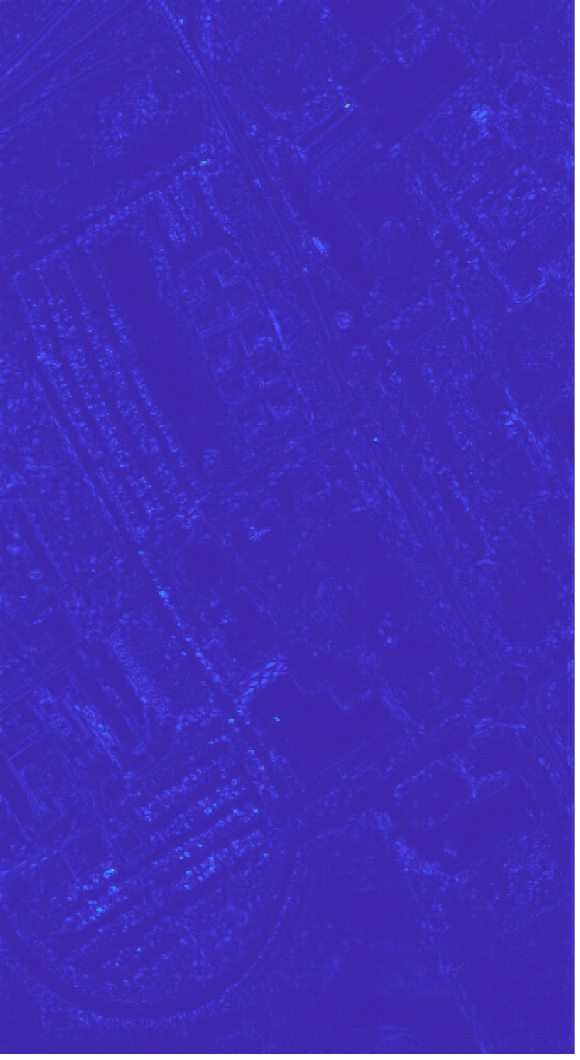} &
   \includegraphics[width=\subfigwidth]{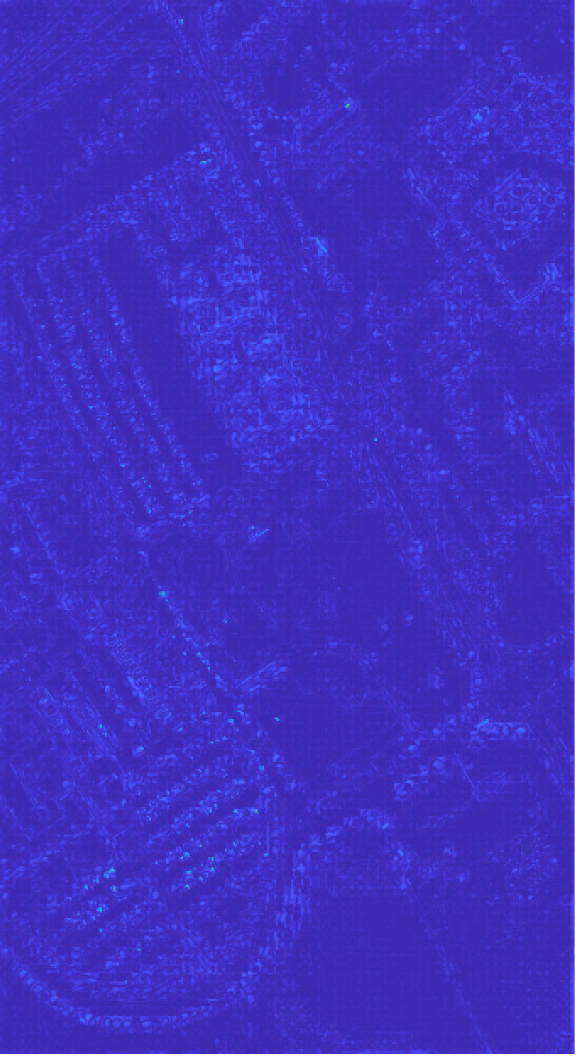} &
   \includegraphics[width=\subfigwidth,height=3.2cm]{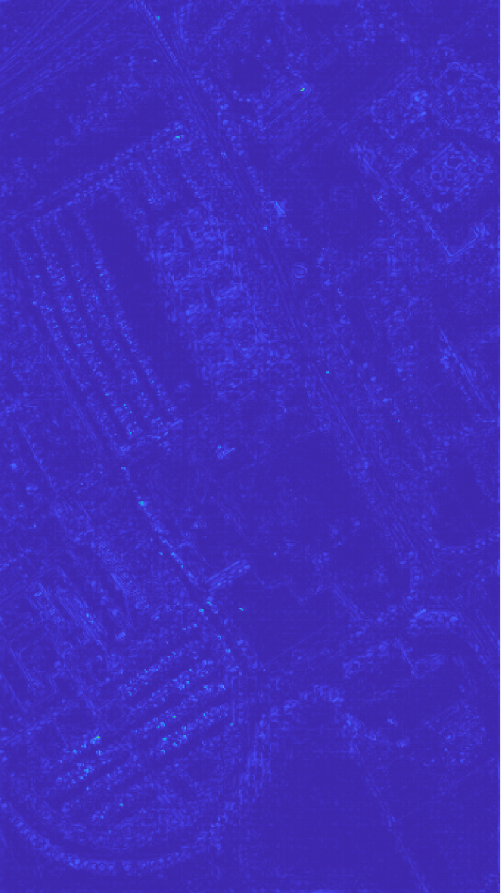} &
   \includegraphics[width=\subfigwidth,height=3.2cm]{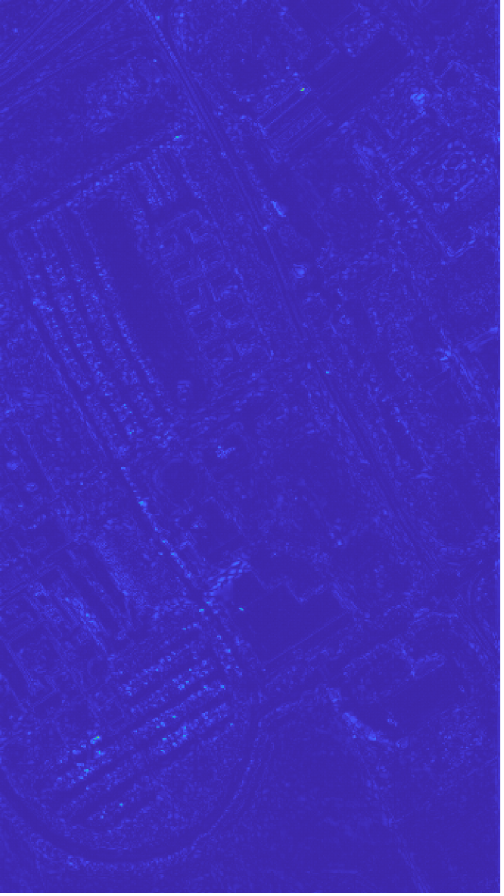} &
   \includegraphics[width=\subfigwidth,height=3.2cm]{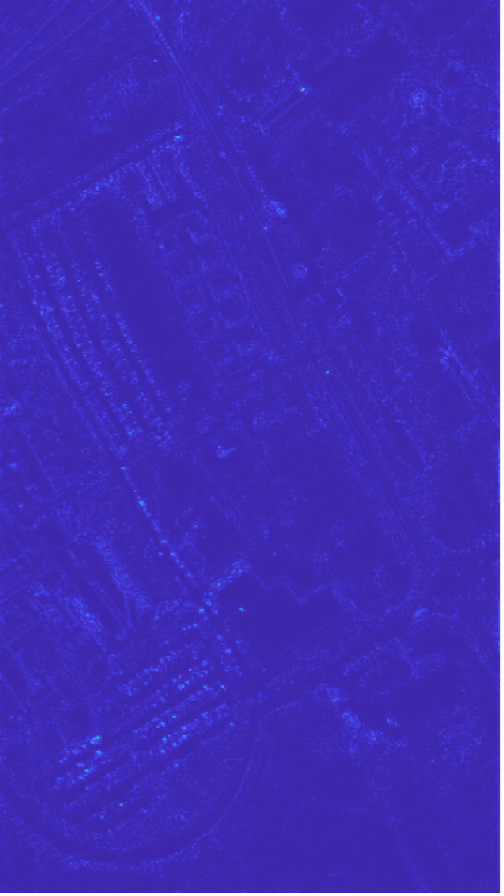}
\end{tabular}
}
  \caption{Fusion experiment with the Pavia University data set -- Color compositions of the fused image (1st row) and corresponding error images (2nd row).}\label{fig_pavia_map}
\end{figure*}


\setlength{\tabcolsep}{1pt}
\begin{figure*}
  \centering
  \scriptsize
  \resizebox{\textwidth}{!}{
  \begin{tabular}{cccccccccc}
  HySure & FUSE-S & CMS & Deep-HS-prior & CNN-Fus & GDD & Adam-VAE & ADMM-VAE  & Adam-GDD & ADMM-GDD\\
  \includegraphics[width=\subfigwidth]{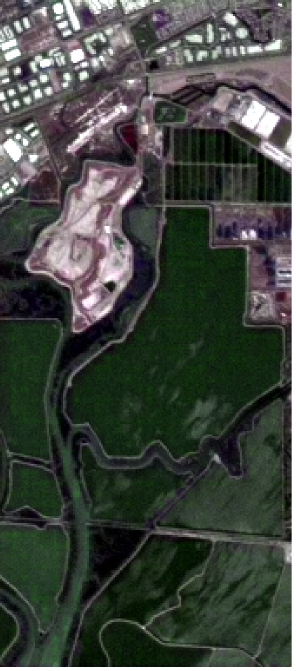} &
  \includegraphics[width=\subfigwidth]{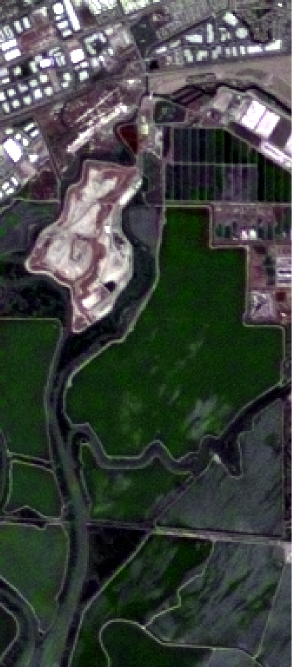} &
  \includegraphics[width=\subfigwidth]{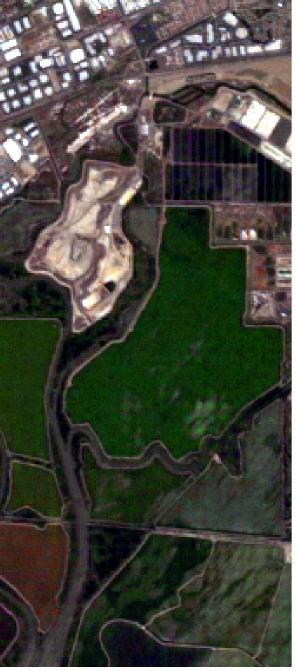} &
  \includegraphics[width=\subfigwidth]{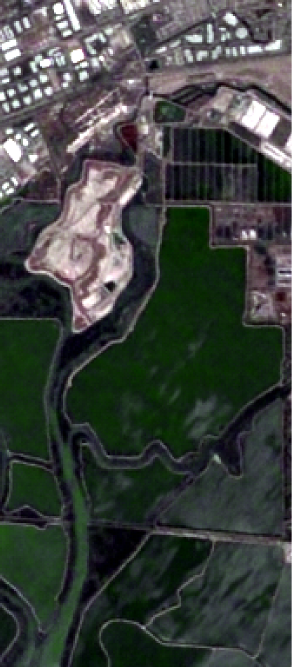} &
  \includegraphics[width=\subfigwidth]{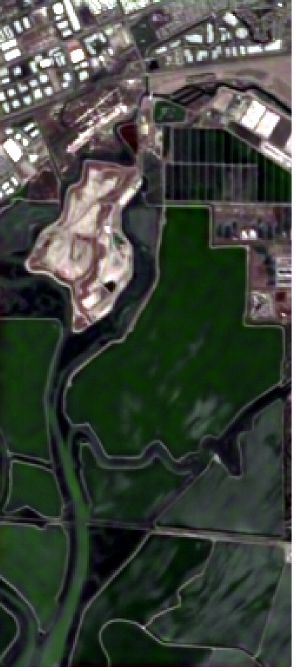} &
  \includegraphics[width=\subfigwidth]{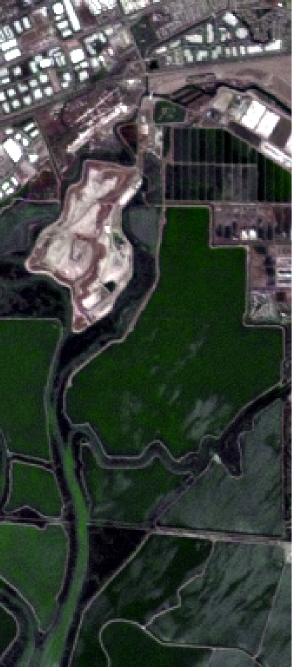} &
  \includegraphics[width=\subfigwidth]{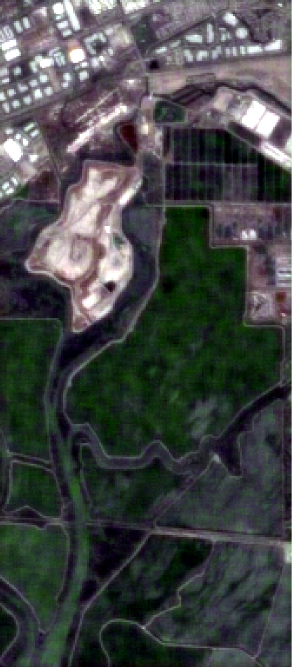} &
  \includegraphics[width=\subfigwidth]{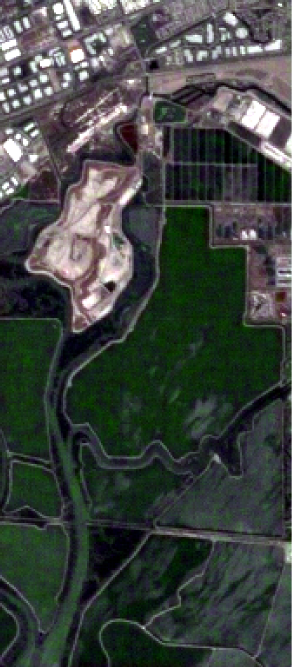} &
  \includegraphics[width=\subfigwidth]{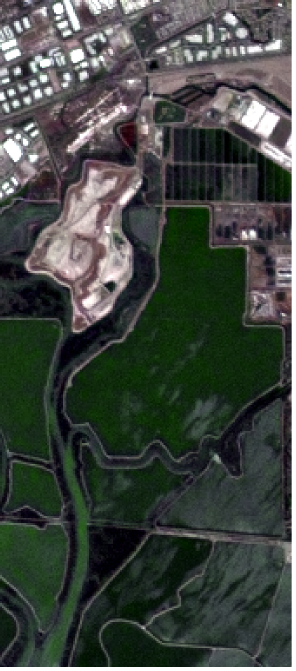} &
  \includegraphics[width=\subfigwidth]{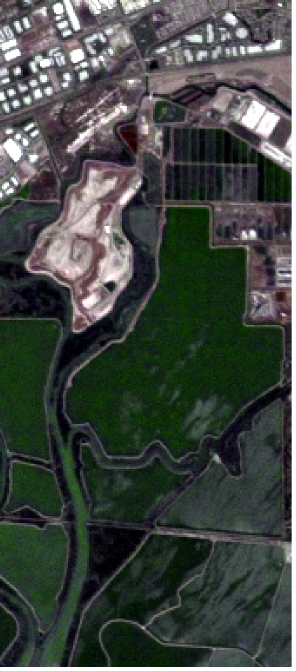}
  \\
  \includegraphics[width=\subfigwidth]{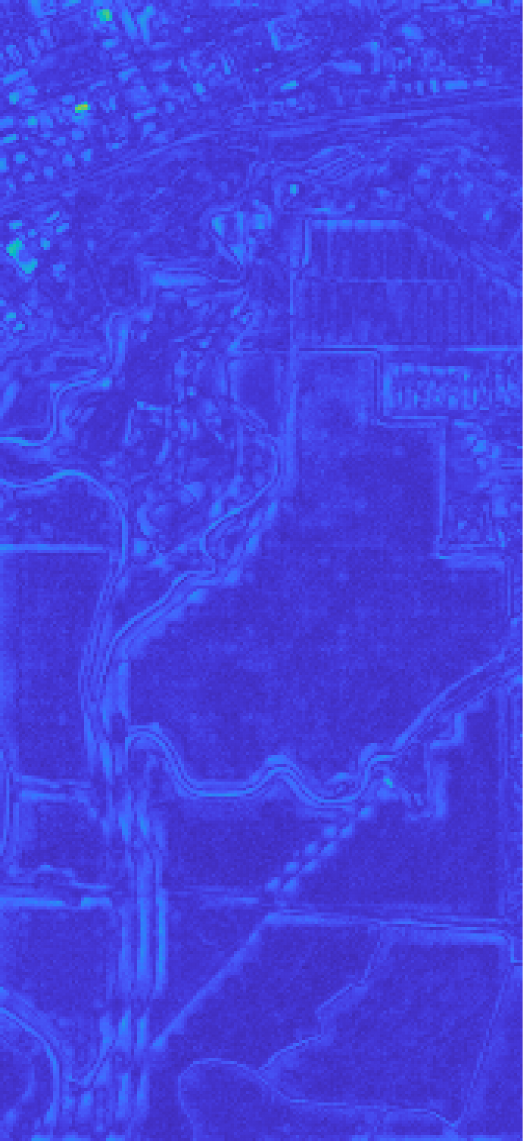} &
  \includegraphics[width=\subfigwidth]{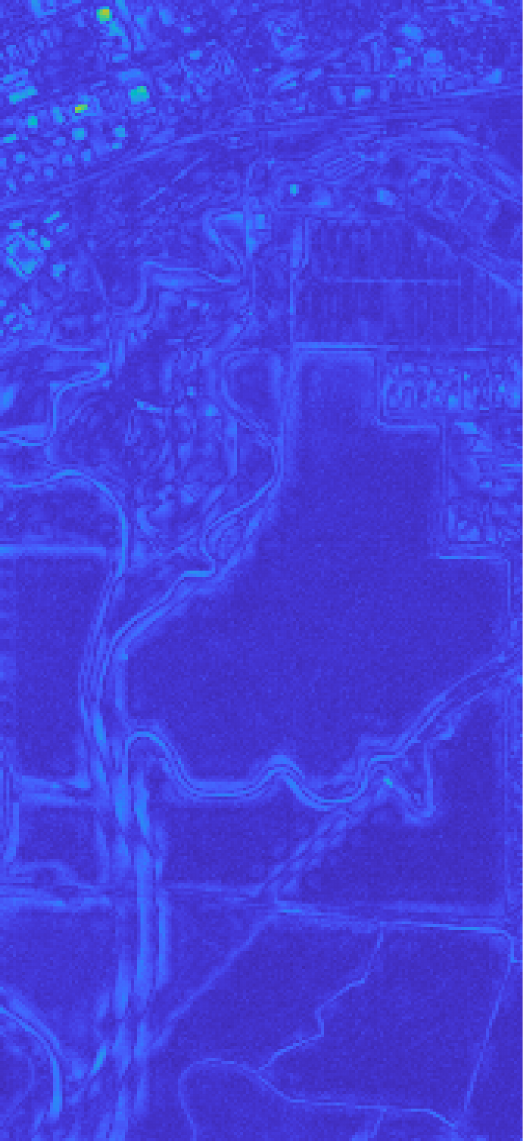} &
  \includegraphics[width=\subfigwidth]{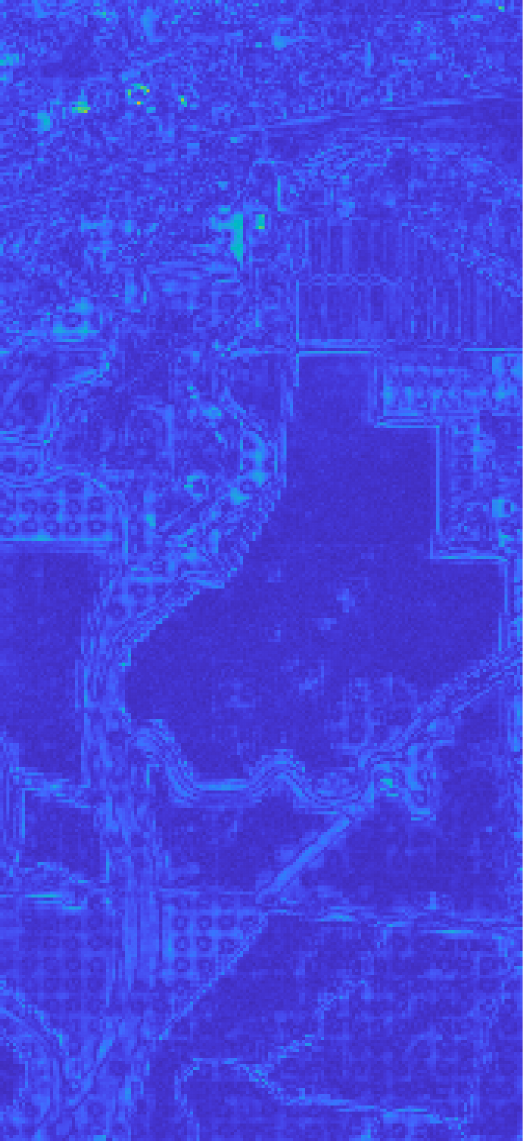} &
  \includegraphics[width=\subfigwidth]{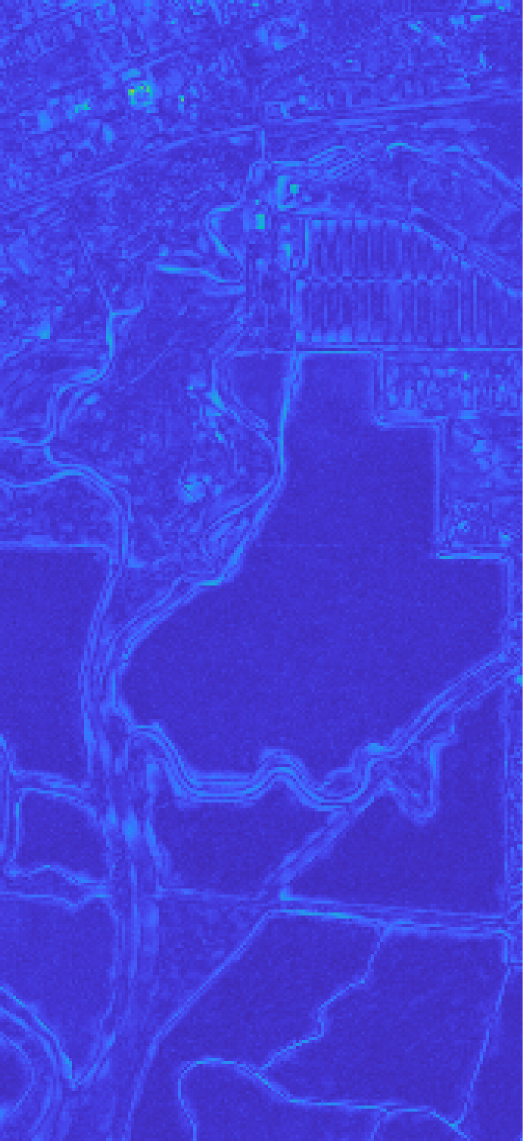} &
  \includegraphics[width=\subfigwidth]{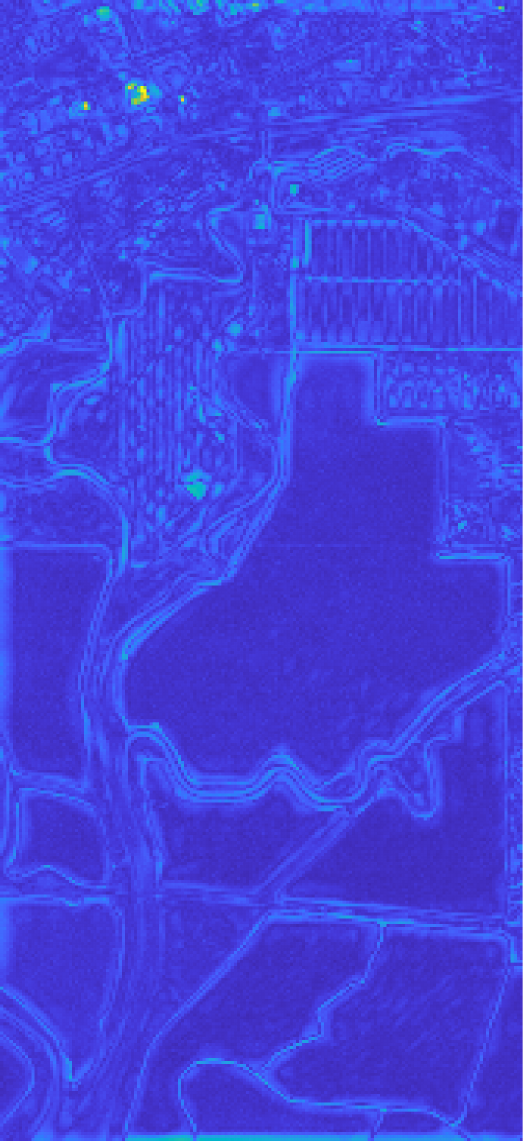} &
  \includegraphics[width=\subfigwidth]{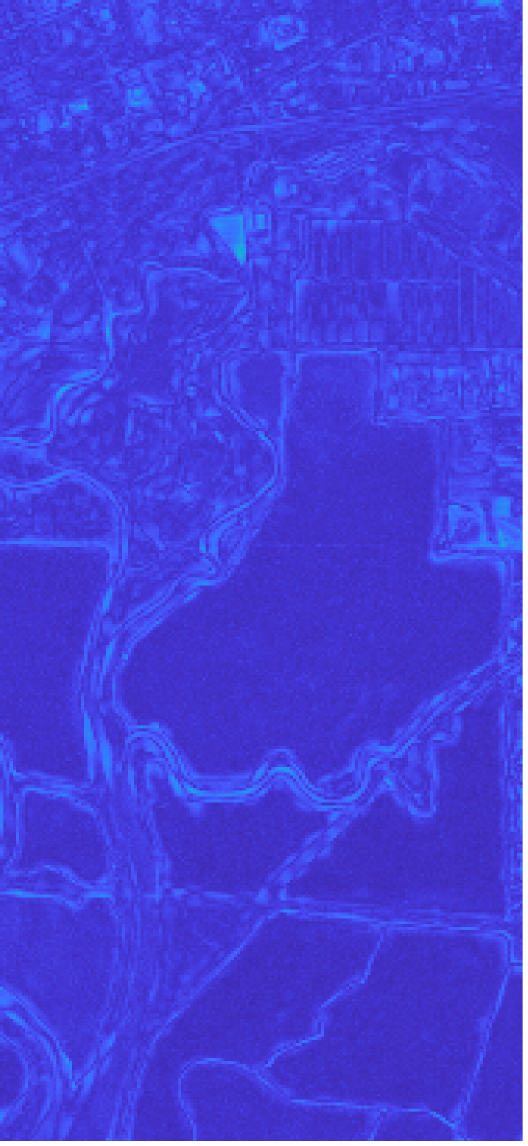} &
  \includegraphics[width=\subfigwidth]{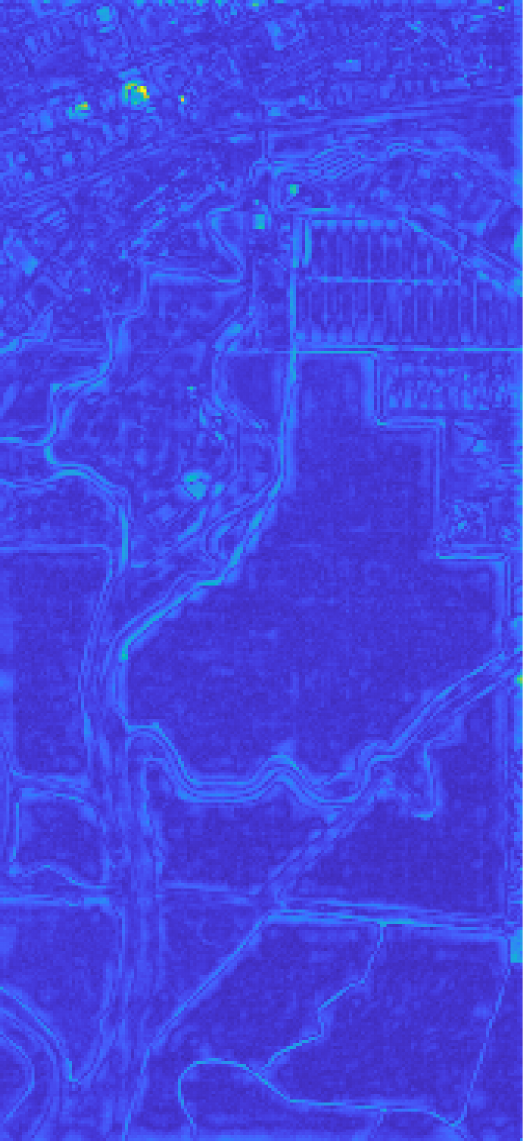} &
\includegraphics[width=\subfigwidth]{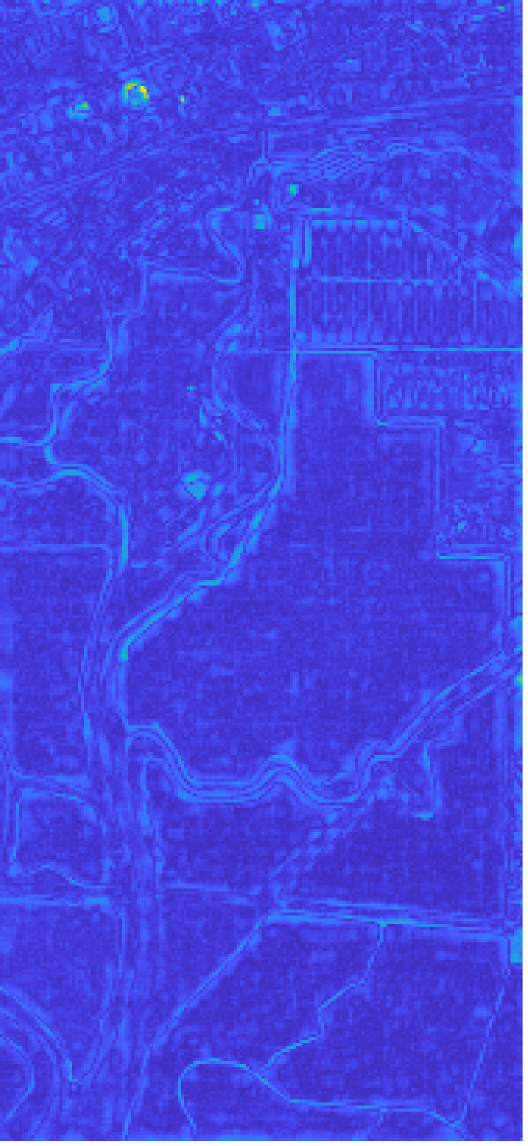} &
  \includegraphics[width=\subfigwidth]{fig/moffett_fusion_results/moffett_error_ADAM_VAE.pdf} &
  \includegraphics[width=\subfigwidth,height=3.8cm]{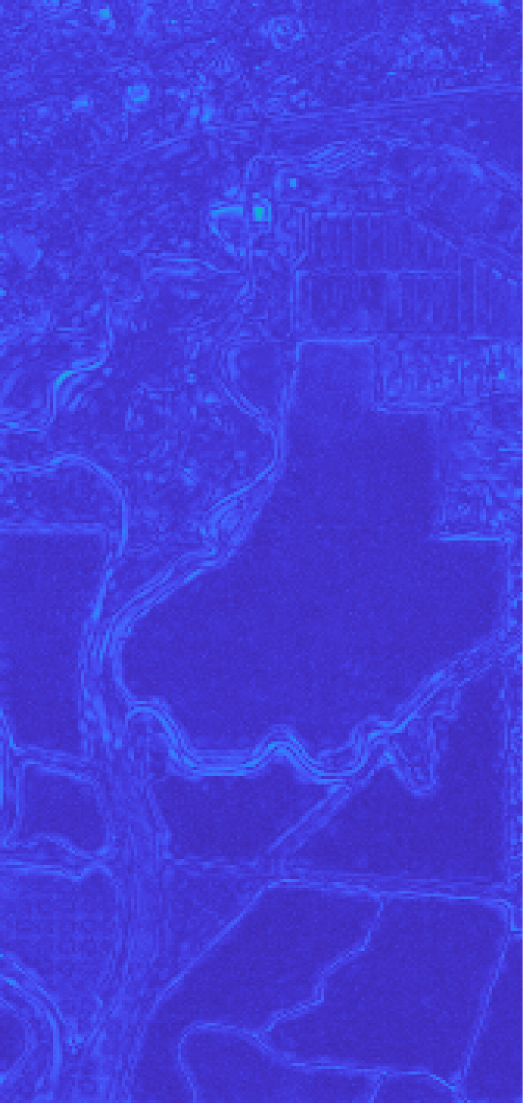}
\end{tabular}
}
  \caption{Fusion experiment with the Moffett field image data set -- Color composition of the fused image (1st row) and corresponding error images (2nd row).}\label{fig_moffet_map}
\end{figure*}

\begin{table}
\scriptsize
\centering
\renewcommand\arraystretch{1.5}
\caption{Fusion experiment with the Pavia University data set -- Quantitative results.}\label{Tab_pavia_results}
\begin{tabular}{c|c|c|c|c|c}
\hline\hline
Methods         & PSNR $\uparrow$  & SAM $\downarrow$ & UIQI $\uparrow$ & ERGAS $\downarrow$ & SSIM $\uparrow$ \\ \hline
HySure          & 29.9739          & 6.9149           & 0.8003          & 4.6984             & 0.7739          \\ \hline
Fuse-S          & 32.2769          & 5.4481           & 0.9302          & 3.2144             & 0.8965          \\ \hline
CMS             & 30.1149          & 7.6407           & 0.8809          & 4.0991             & 0.8239          \\ \hline
Deep-HS-prior   & 32.0470           & 5.3685           & 0.9228          & 3.1756             & 0.8750          \\ \hline
CNN-Fus         & 32.0671           & 5.7449          & 0.9302          & 3.2892             & 0.8940          \\ \hline
GDD         & 33.3612  & 5.1431  & 0.9480   & 2.6877  & 0.9229          \\ \hline
Adam-VAE &29.7176	&5.4624	&0.8688	&4.3543	&0.8077
 \\ \hline
ADMM-VAE &31.5442	&5.3400	&0.8905	&4.1718	&0.8521
 \\ \hline
Adam-GDD       &33.8621	&4.9337	&0.9481	&2.6865	&0.9228
     \\ \hline
ADMM-GDD& \textbf{34.3931} & \textbf{4.4825}  & \textbf{0.9544} & \textbf{2.5635}    & \textbf{0.9276} \\ \hline
\hline
\end{tabular}
\end{table}

\begin{table}
\scriptsize
\centering
\renewcommand\arraystretch{1.5}
\caption{Fusion experiment with the Moffett Field data set -- Quantitative results.}\label{Tab_moffett_results}
\begin{tabular}{c|c|c|c|c|c}
\hline\hline
Methods         & PSNR $\uparrow$  & SAM $\downarrow$ & UIQI $\uparrow$ & ERGAS $\downarrow$ & SSIM $\uparrow$ \\ \hline
HySure          & 33.3613          & 5.1800           & 0.8196          & 4.2658             & 0.7922          \\ \hline
Fuse-S          & 34.4079          & 4.9586           & 0.8295          & 4.1692             & 0.8051          \\ \hline
CMS             & 31.4534          & 5.2291           & 0.7570          & 4.8516             & 0.7183          \\ \hline
Deep-HS-prior   & 33.4047          & 4.9510           & 0.7908          & 4.3346             & 0.7598          \\ \hline
CNN-Fus         & 33.2445          & 5.0889           & 0.7653          & 4.8852             & 0.7257          \\ \hline
GDD  &34.6076	&4.8237	 &0.8468	&3.9714	&0.8252\\ \hline
Adam-VAE &31.4815 &5.7846	&0.7485	&5.8547	&0.6675
 \\ \hline
ADMM-VAE &33.5162	&5.0414	&0.8053	&4.8742	&0.7899
 \\ \hline
Adam-GDD  &34.7201	&4.7894	&0.8495	&3.9831	&0.8271 \\ \hline
ADMM-GDD&\textbf{34.9603}	&\textbf{4.7782} &\textbf{0.8635}	&\textbf{3.6390}	&\textbf{0.8356} \\ \hline\hline
\end{tabular}
\end{table}

\subsection{Multiband image fusion}\label{subsec:results_fusion}

\noindent \textbf{Data --} In this study, two simulated hyperspectral data sets, namely the Pavia University and Moffett field data sets, are used to evaluate the effectiveness of the proposed method when tackling a multiband image fusion problem (see Section \ref{subsec:app_fusion}).

The Pavia University image was acquired over the urban area of Pavia University, Italy. It consists of $610\times340$ pixels ($N=207400$) with $B=93$ spectral bands after removing the water vapor absorption and noisy bands. A color composition of the image is depicted in Fig.~\ref{fig_sr_RGB}(a). It is considered as the ground-truth reference image $\mathbf{X}$ of high spatial and high spectral resolutions to be recovered. The observed images have been synthetically generated from this reference image following a protocol similar to the one described in \cite{wei2015fast}. More precisely, a hyperspectral image $\mathbf{Y}_{\mathrm{HS}}$ of low spatial resolution is generated by applying a $5\times5$ Gaussian filter with a standard deviation set to 2 and then subsampling with a factor equal to 5 in the horizontal and vertical directions. A panchromatic image $\mathbf{Y}_{\mathrm{HR}}$ of high spatial resolution is obtained from the reference image by averaging all bands. Zero-mean additive Gaussian noises are added to the observed images with corresponding noise levels of SNR$= 35$dB for the hyperspectral image and SNR$ = $30dB for the panchromatic image.

The Moffett field image was acquired by the JPL/NASA airborne visible/infrared imaging spectrometer (AVIRIS). This reference image is of size $396\times184$ pixels with $B=176$ spectral bands after removing the water vapor absorption and noisy bands. Fig.~\ref{fig_sr_RGB}(b) depicts a color composition of the image. As for the previous data set. To produce a hyperspectral image $\mathbf{Y}_{\mathrm{HS}}$ of lower spatial resolution, the reference image $\mathbf{X}$ has been spatially degraded by applying a
$7\times7$ Gaussian filter with standard deviation set to 2 and then subsampling with a factor set to 7 in both directions. Conversely, a panchromatic image $\mathbf{Y}_{\mathrm{HR}}$ has been generated by averaging the visible bands (1-41 bands). The two images have been corrupted by zero-mean additive Gaussian noises with SNR$ = 30$dB for the hyperspectral image $\mathbf{Y}_{\mathrm{HS}}$ and SNR$= 35$dB for the panchromatic image $\mathbf{Y}_{\mathrm{HR}}$.\\

\noindent \textbf{Compared methods --} The proposed method have been compared with some recently proposed fusion methods, including HySure~\cite{simoes2014convex}, FUSE-S~\cite{wei2015hyperspectral}, CMS~\cite{zhang2018exploiting}, Deep-HS-prior~\cite{sidorov2019deep} and CNN-Fus~\cite{dian2020regularizing}. The guided deep decoder (GDD) used to define the generative model in the proposed framework is also considered as an end-to-end fusion method~\cite{uezato2020guided}. For the HySure algorithm which relies on a total variation regularizer to enforce the smoothness of the fused image, the hyperparameters have been set to $\lambda_{\rho}= 1\times10^{-3}$ and $\lambda_{m}= 10$. The FUSE-S method exploits a sparse representation as a regularization and sparsity parameter $\lambda_{s}$ is set to 25. The CMS method excavates clustering manifold structure for super-resolution task and the corresponding hyperparameters have been set to  $\mu_{\text{CMS}}=4\times10^{-4}$ and $\rho_{\text{CMS}}= 1.05$. Deep-HS-prior is a hyperspectral image enhancement method based on a deep image prior architecture. To train this model, the cost function is defined to ensure consistency of the pair of the hyperspectral image of low spatial resolution and the panchromatic image of high spatial resolution. The CNN-Fus is a plug-and-play based hyperspectral super-resolution method where a CNN-denoiser acts as a regularization. To train the VAE and GDD models, the number of epochs is set to 100 and 7000 and the learning rates of the Adam optimizer are fixed to $1\times 10^{-3}$ and $0.01$, respectively. To train the VAE model, 5000 patches have been extracted from the guidance image and 10000 patches from an external image data set.
The parameters of the proposed algorithm are set as $\mu=1\times10^{-4}$ and $\lambda=1\times10^{-5}$.\\

\noindent \textbf{Results --} The quantitative results associated with the Pavia University and Moffett field data sets are reported in Table~\ref{Tab_pavia_results} and \ref{Tab_moffett_results}, respectively, where the best results are highlighted in bold. Several important findings can be drawn for these results. Firstly, GDD competes favorably with respect to the three compared model-based algorithms (HySure, FUSE-S, CMS) and to the data-driven algorithms Deep-HS-prior and CNN-Fus. These good results can be explained by the relevant deep architecture able to extract meaningful spatial information to guide the inversion process. In addition, when using the decoder as a guided generative model employed as a regularization, it provides even better performance. Indeed, the results obtained by Adam-GDD and ADMM-GDD demonstrates the relevance of the designed objective function \eqref{eq:optim2_1}. Secondly, the splitting-based minimization scheme ADMM-GDD proposed in Section \ref{subsec:optimization} provides significantly better results compared to a direct minimization by the Adam solver (Adam-GDD). This is also observed when the GDD-based regularization is replaced by a VAE model: ADMM-VAE performs better than Adam-VAE. Finally, the GDD regularization is shown to better capture the spatial information when compared to the VAE regularization, whatever the minimization technique (Adam or ADMM). This can be explained by the fact that GDD is guided by the sole guidance image of high spatial resolution while the VAE has been trained on an extended data set. In conclusion, the proposed ADMM-GDD method outperforms all compared methods since it combines the advantages of GDD as the regularization and ADMM as the minimization scheme. Fig.~\ref{fig_pavia_map} and Fig.~\ref{fig_moffet_map} illustrate the fusion results by depicting color compositions of the fused image recovered by the compared algorithms. It also depicts the spatial map of the pixelwise reconstructed errors averaged over the spectral bands. It can be observed that the proposed method reconstructs more details and preserves their sharpness.

\newcommand{\subfigwidths}{1.45cm}
\setlength{\tabcolsep}{0.8pt}
\begin{figure*}
  \centering
  \scriptsize
  \resizebox{\textwidth}{!}{
  \begin{tabular}{cccccccccccc}
 Reference & Measured & FastHyIn & Deep-HS-prior & WLRTR & PnP-In & ADAMADMM & GDD & Adam-VAE & ADMM-VAE & Adam-GDD & ADMM-GDD\\
   \includegraphics[width=\subfigwidths]{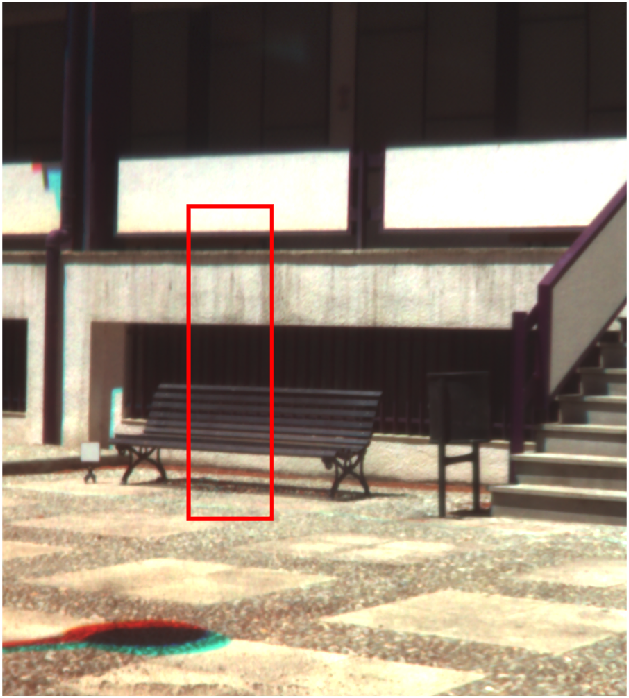} &
   \includegraphics[width=\subfigwidths]{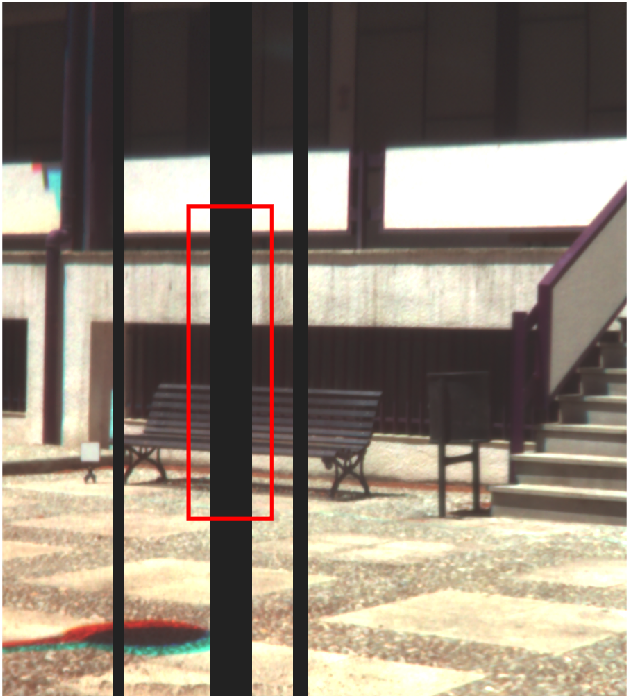} &
   \includegraphics[width=\subfigwidths]{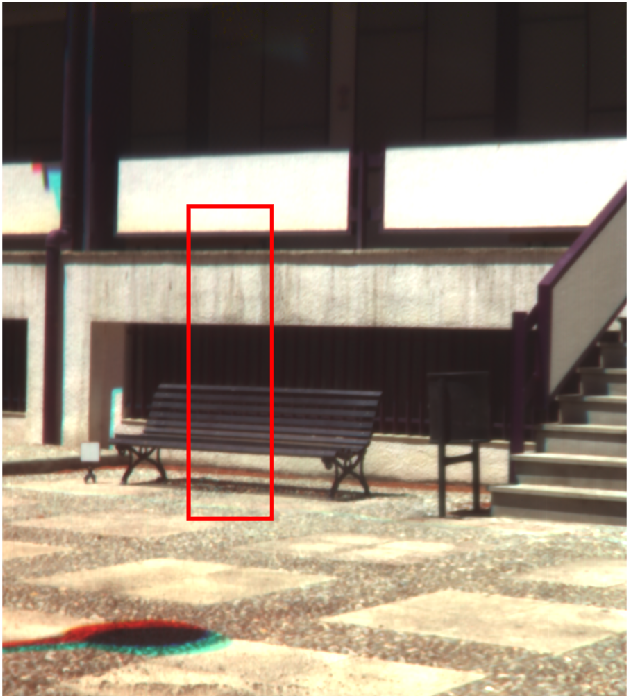} &
   \includegraphics[width=\subfigwidths]{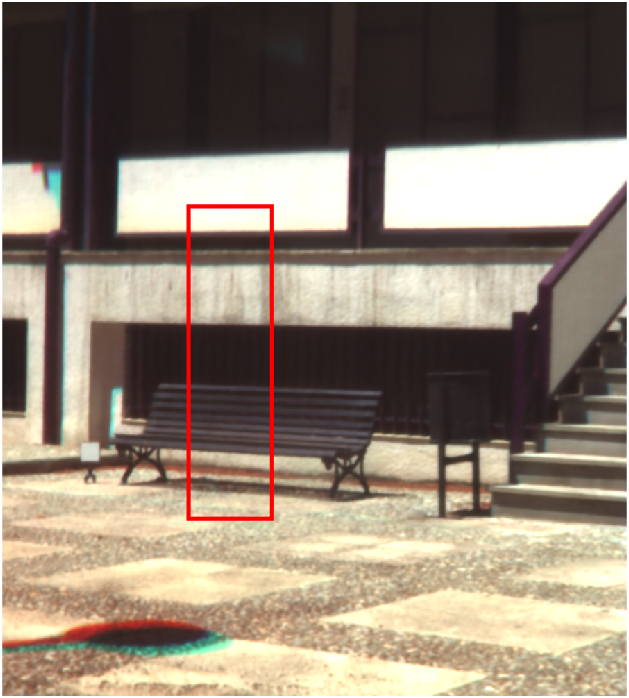} &
   \includegraphics[width=\subfigwidths]{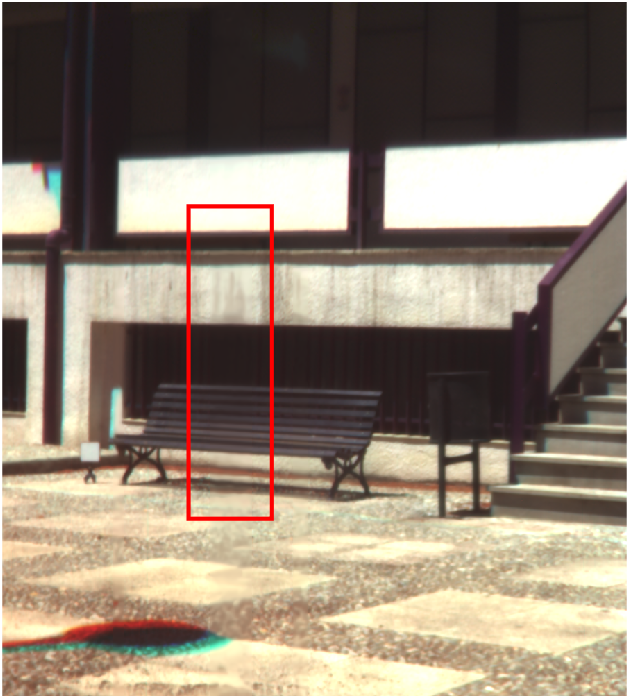} &
   \includegraphics[width=\subfigwidths]{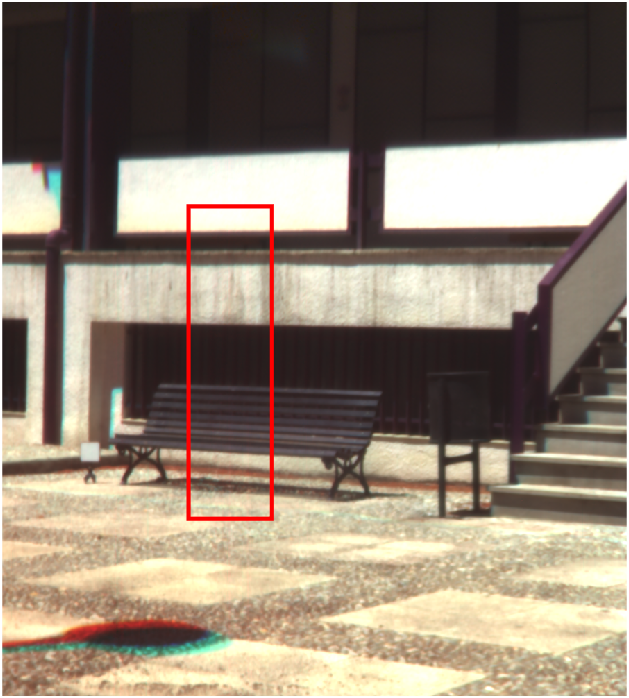} &
   \includegraphics[width=\subfigwidths]{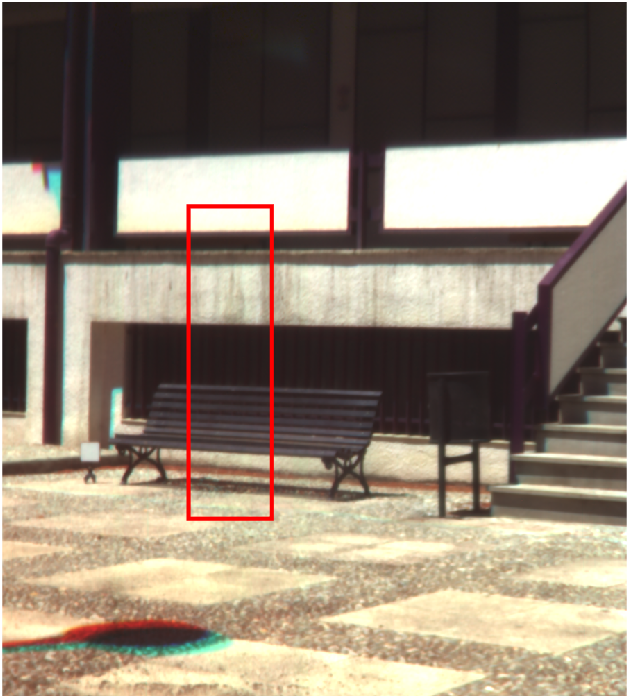} &
   \includegraphics[width=\subfigwidths]{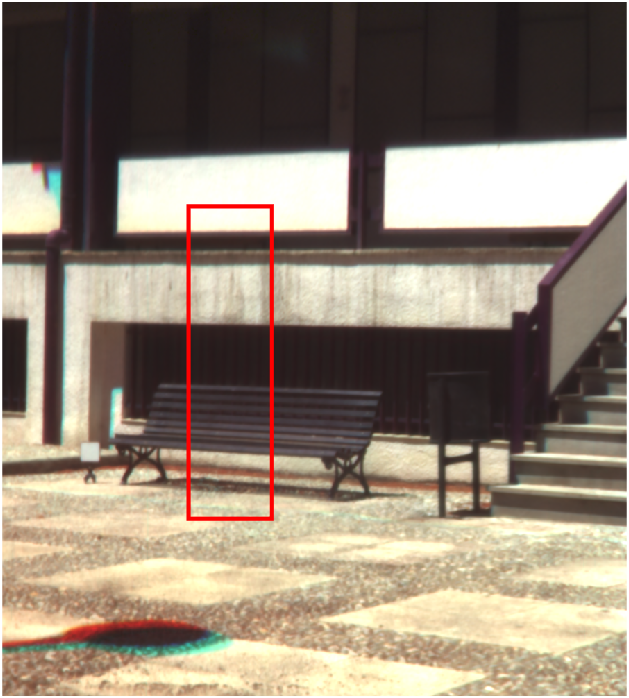} &
   \includegraphics[width=\subfigwidths]{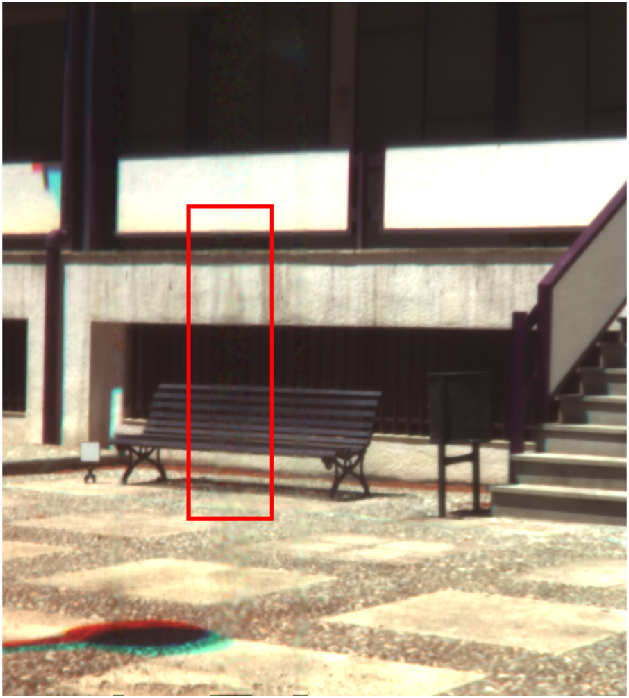} &
   \includegraphics[width=\subfigwidths]{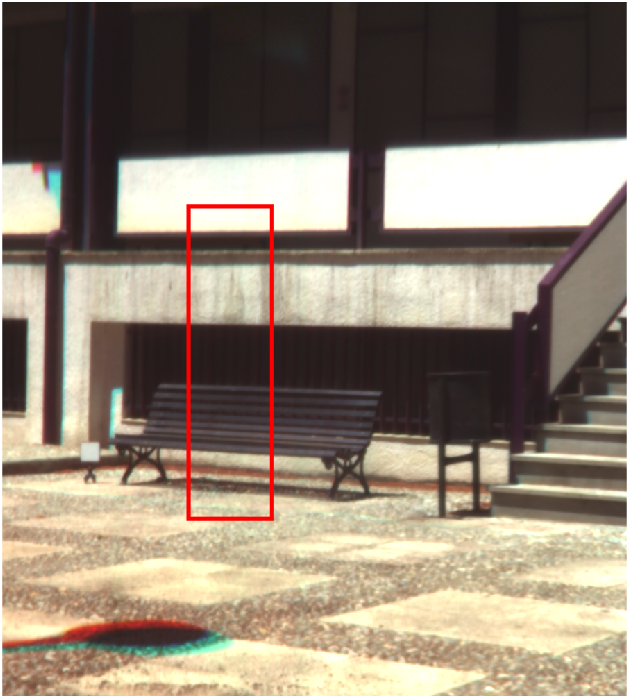}&
   \includegraphics[width=\subfigwidths]{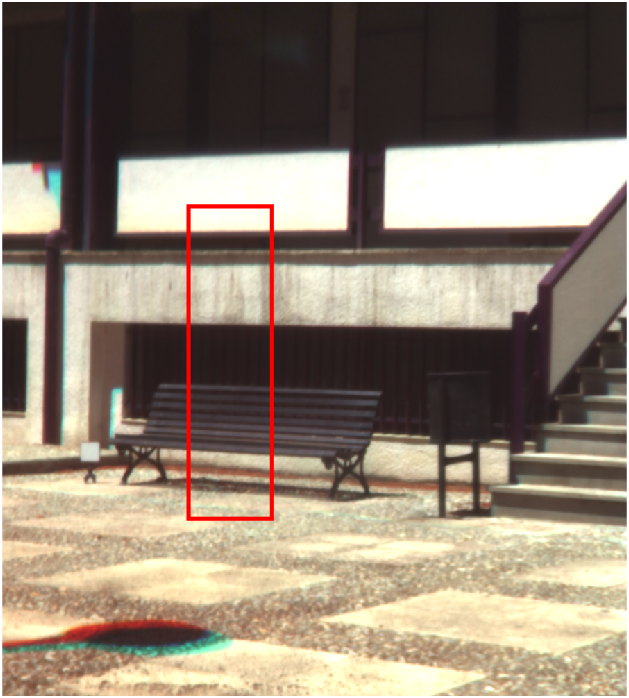}&
   \includegraphics[width=\subfigwidths]{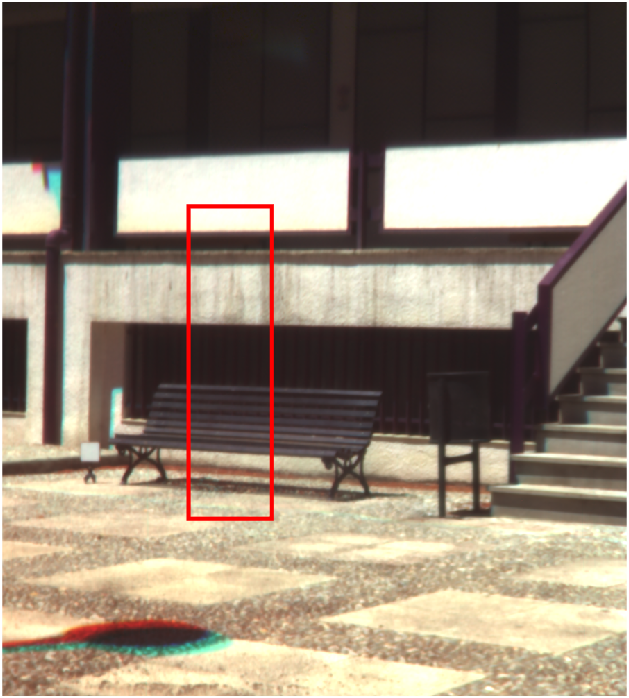}
   \\
   \includegraphics[width=\subfigwidths]{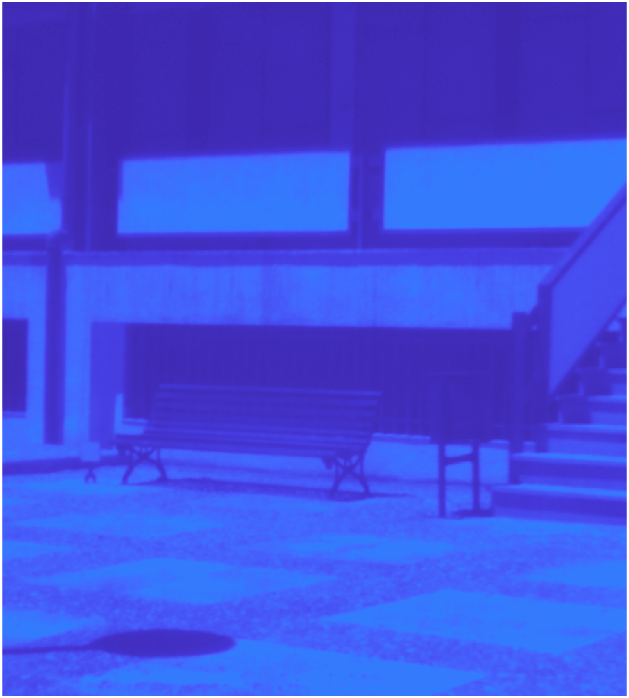} &
   \includegraphics[width=\subfigwidths]{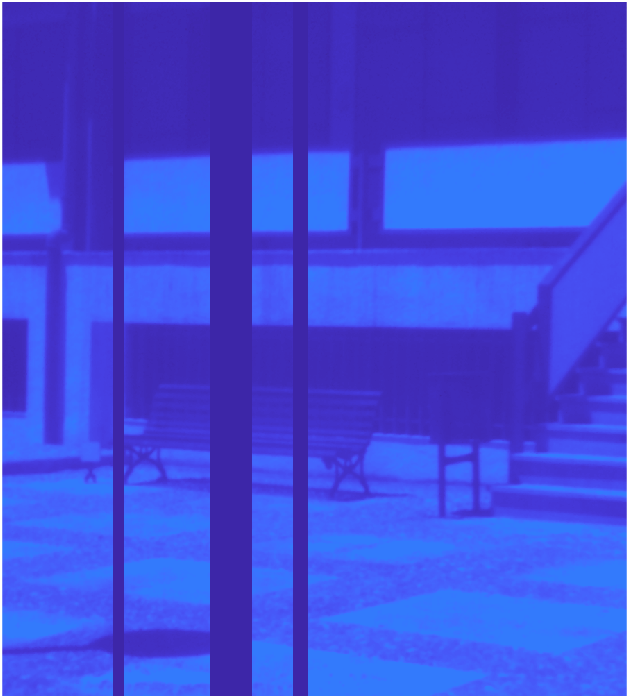} &
   \includegraphics[width=\subfigwidths]{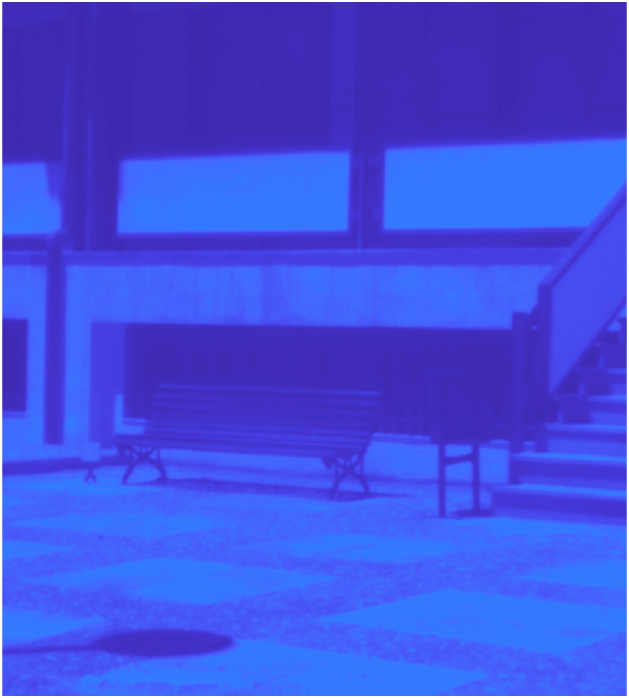} &
   \includegraphics[width=\subfigwidths]{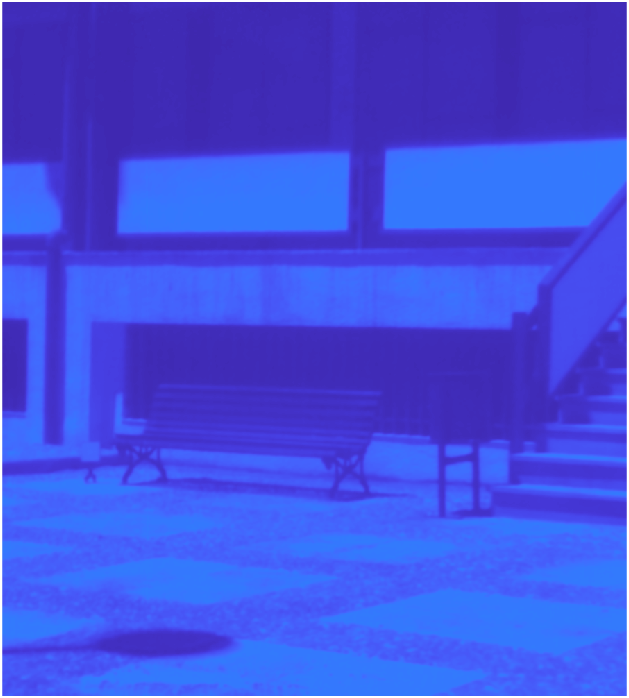} &
   \includegraphics[width=\subfigwidths]{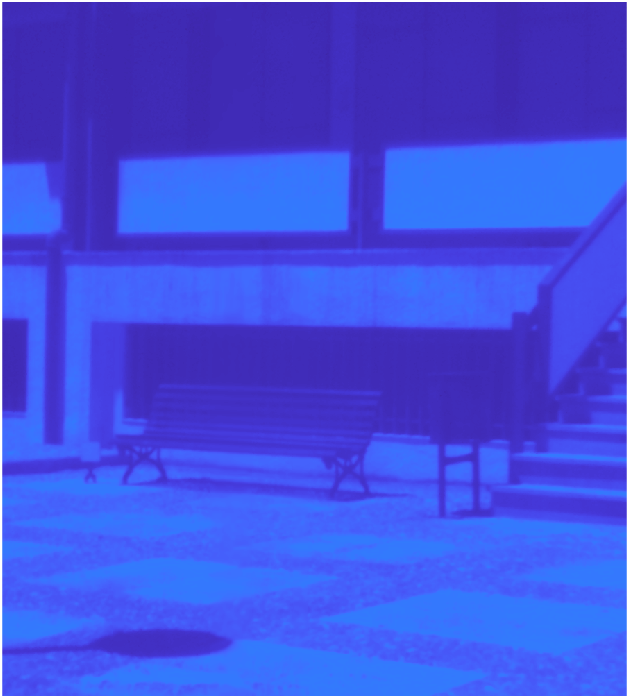} &
   \includegraphics[width=\subfigwidths]{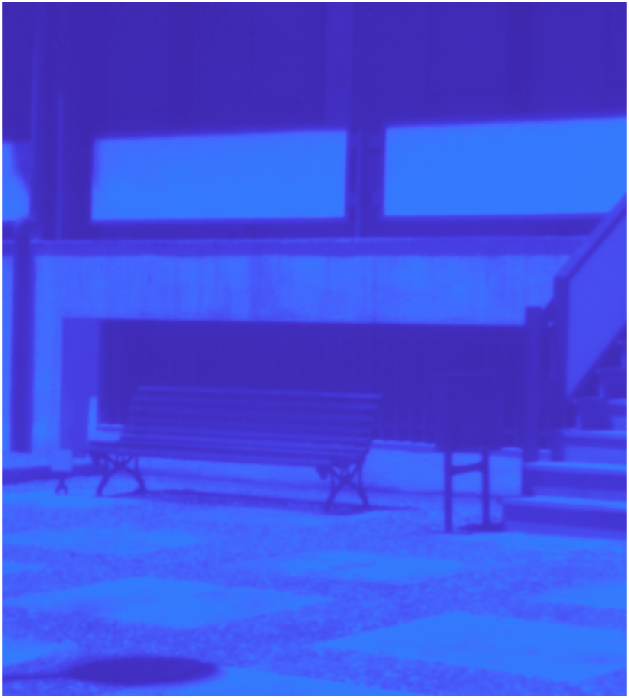} &
   \includegraphics[width=\subfigwidths]{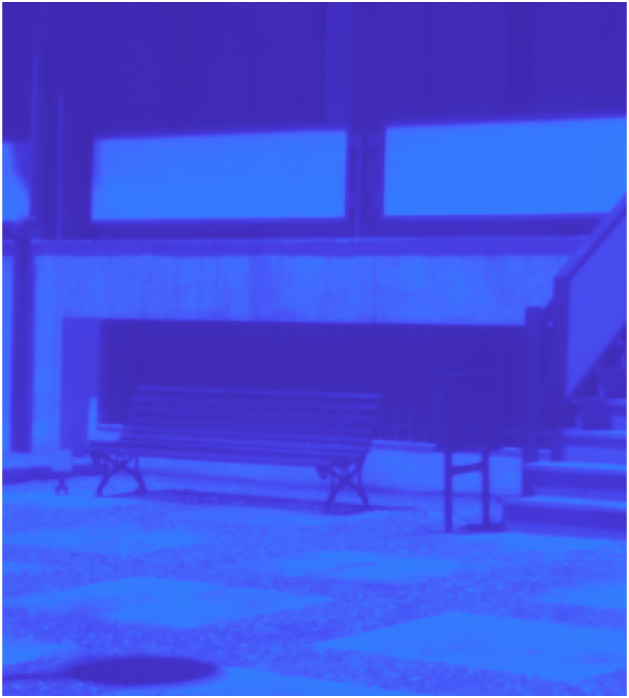} &
   \includegraphics[width=\subfigwidths]{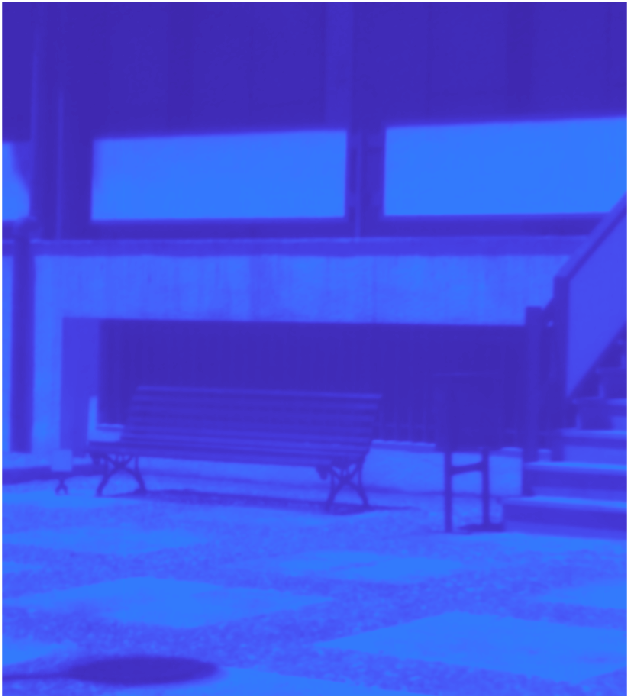} &
   \includegraphics[width=\subfigwidths]{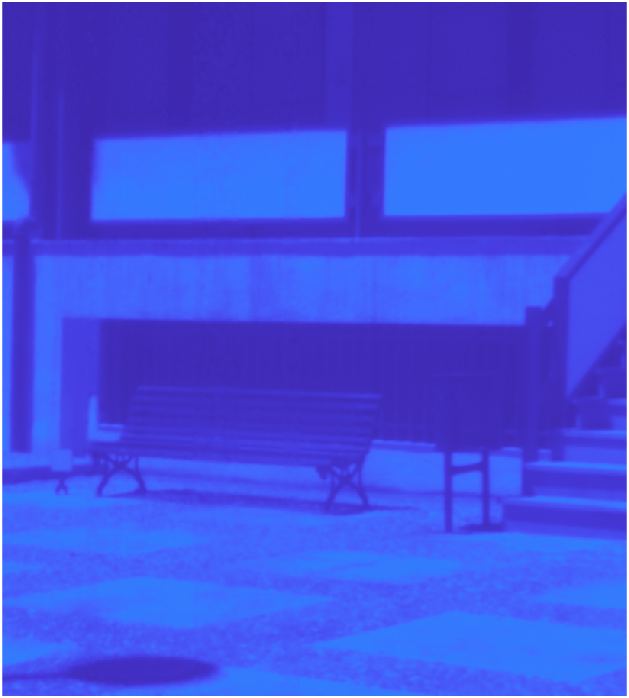} &
   \includegraphics[width=\subfigwidths]{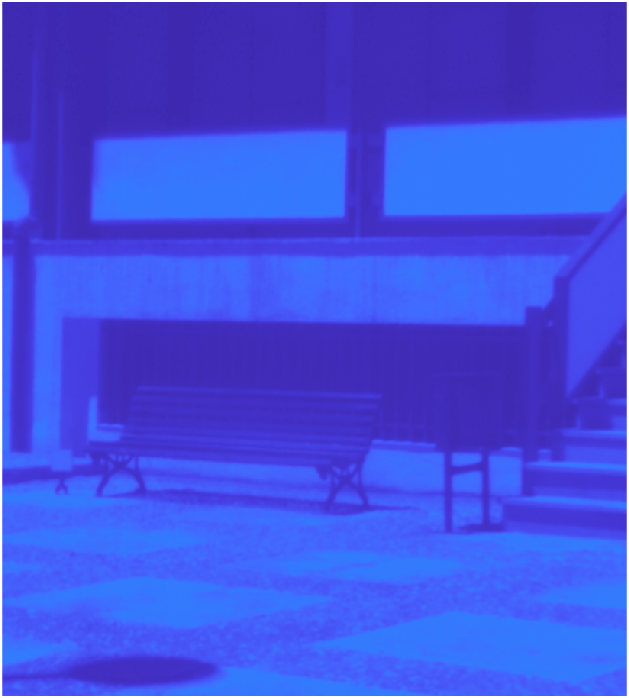} &
   \includegraphics[width=\subfigwidths]{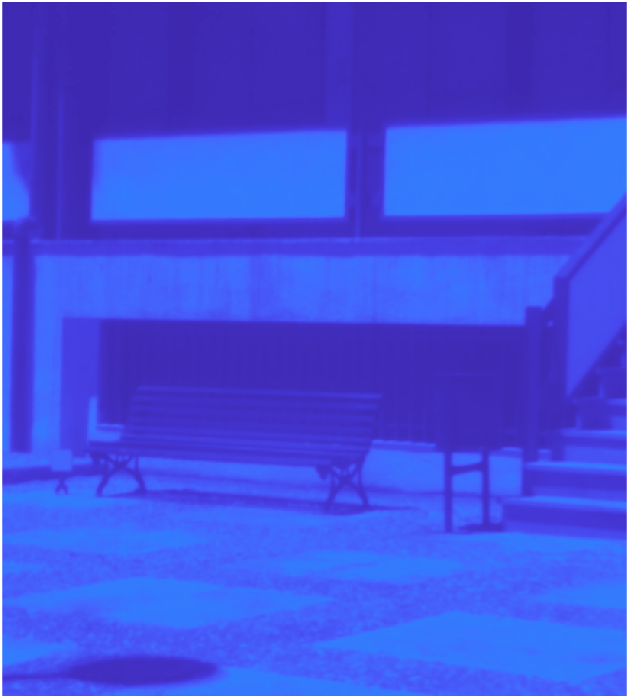} &
   \includegraphics[width=\subfigwidths]{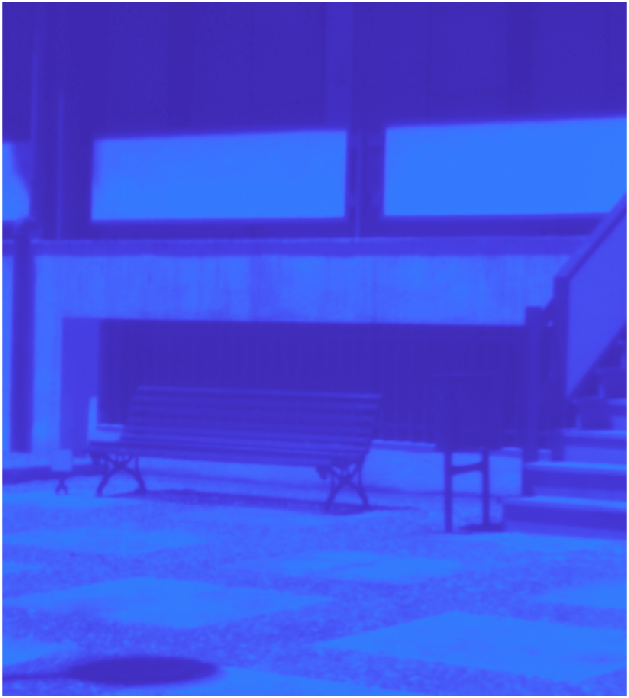}
\end{tabular}}
  \caption{Inpainting experiment with the UGR data set -- Color compositions (1st row) and 47th band (2nd row) of the images. The dead (masked) pixels appear as vertical black lines in the measured images.}\label{fig_scene2_map}
\end{figure*}

\setlength{\tabcolsep}{0.8pt}
\begin{figure*}
  \centering
  \scriptsize
  \resizebox{\textwidth}{!}{
  \begin{tabular}{cccccccccccc}
 Reference & Measured & FastHyIn & Deep-HS-prior & WLRTR & PnP-In & ADAMADMM & GDD & Adam-VAE & ADMM-VAE & Adam-GDD & ADMM-GDD\\
   \includegraphics[width=\subfigwidths]{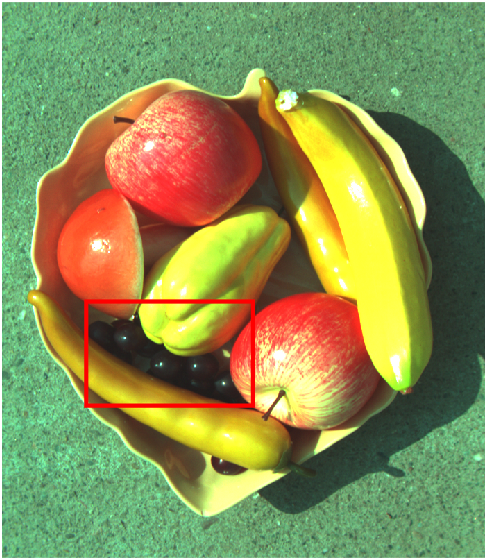} &
   \includegraphics[width=\subfigwidths]{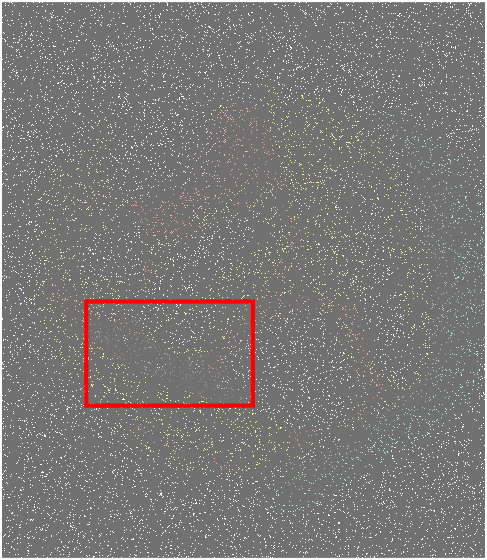} &
   \includegraphics[width=\subfigwidths]{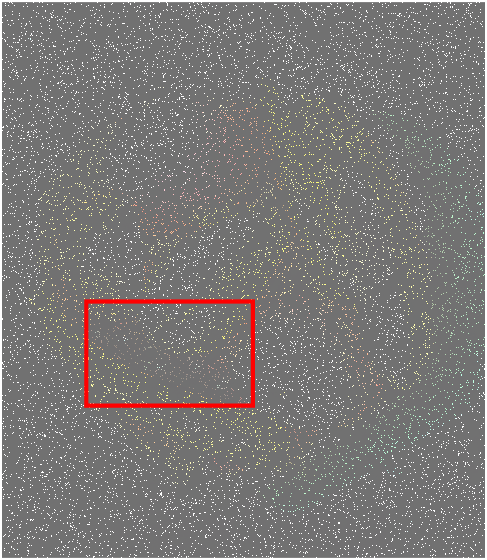} &
   \includegraphics[width=\subfigwidths]{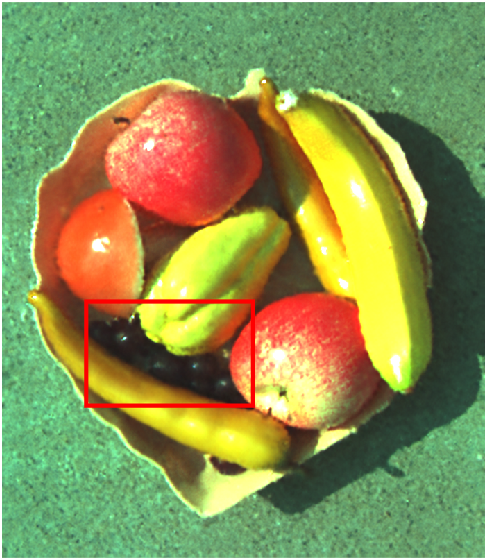} &
   \includegraphics[width=\subfigwidths]{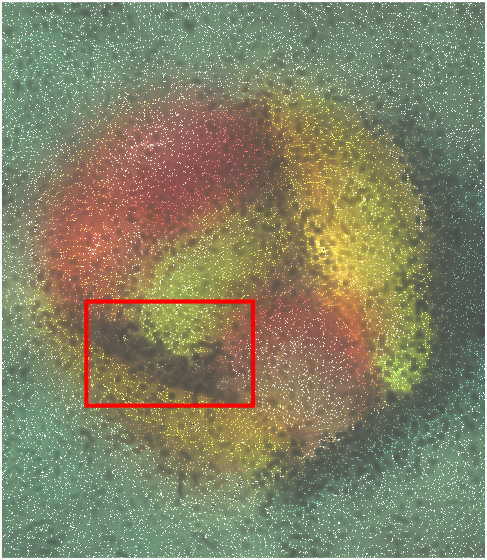} &
   \includegraphics[width=\subfigwidths]{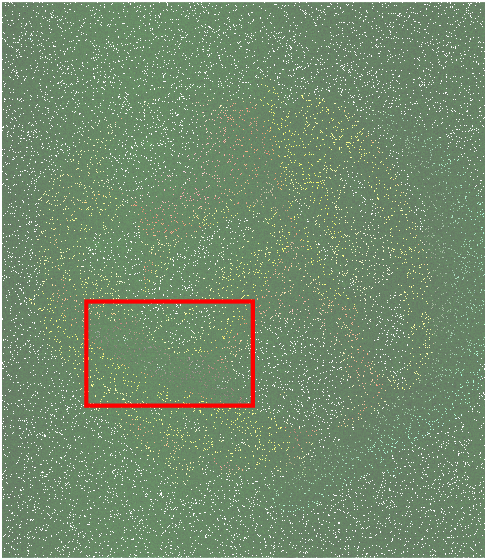} &
   \includegraphics[width=\subfigwidths]{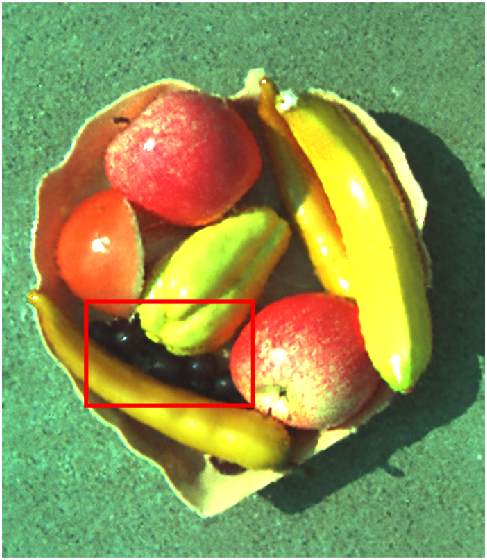} &
   \includegraphics[width=\subfigwidths]{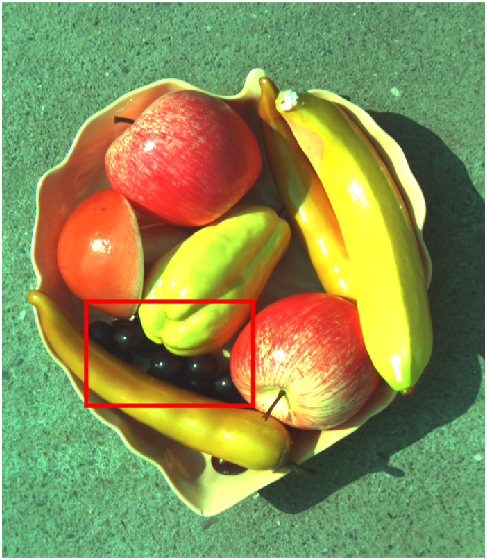} &
   \includegraphics[width=\subfigwidths]{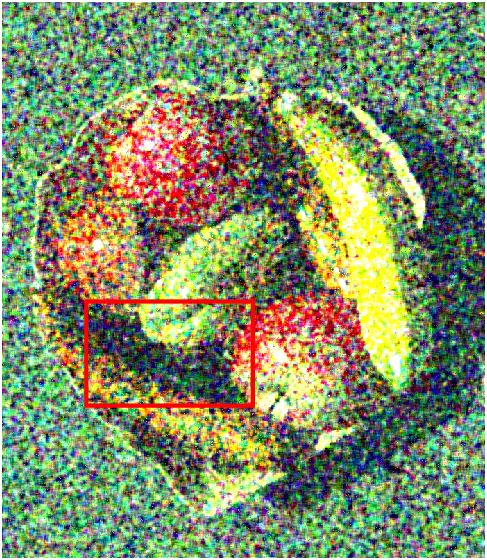} &
   \includegraphics[width=\subfigwidths]{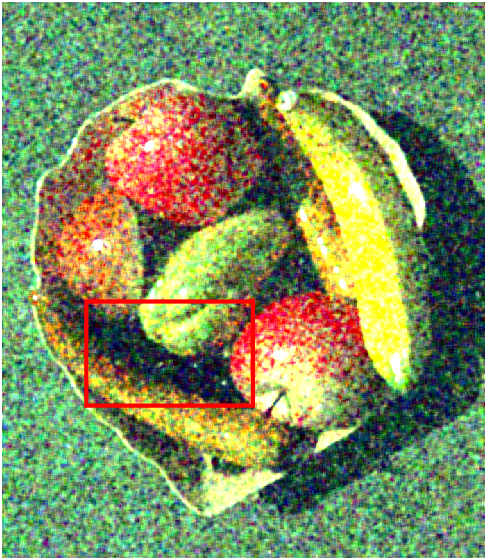}&
   \includegraphics[width=\subfigwidths]{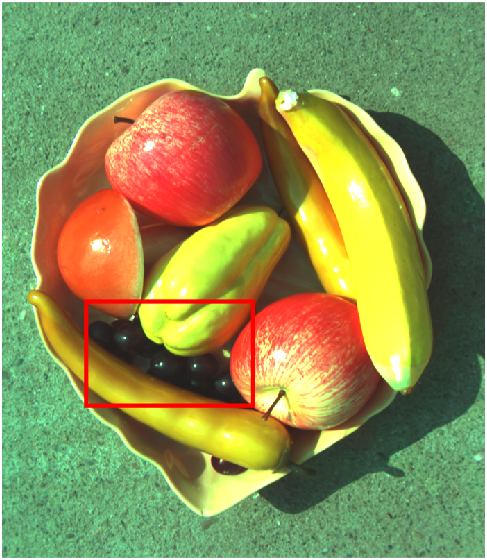}&
   \includegraphics[width=\subfigwidths]{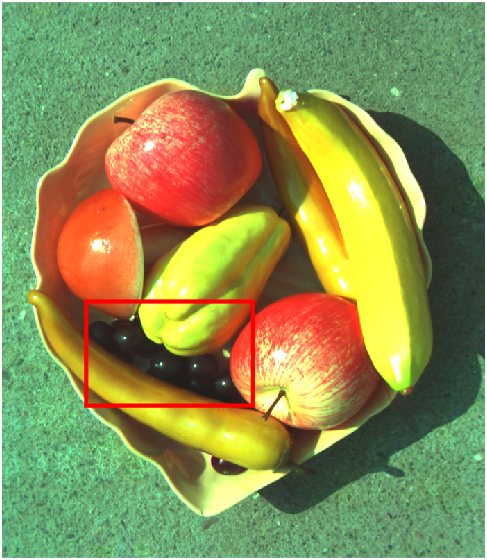}
   \\
   \includegraphics[width=\subfigwidths]{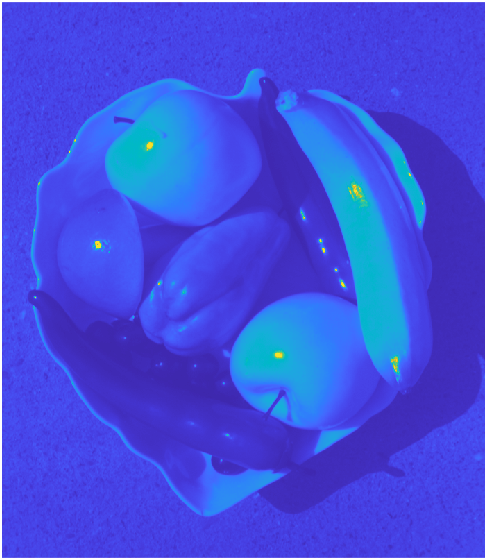} &
   \includegraphics[width=\subfigwidths]{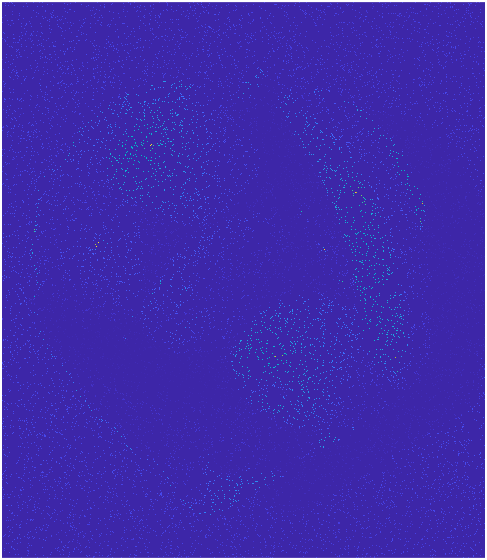} &
   \includegraphics[width=\subfigwidths]{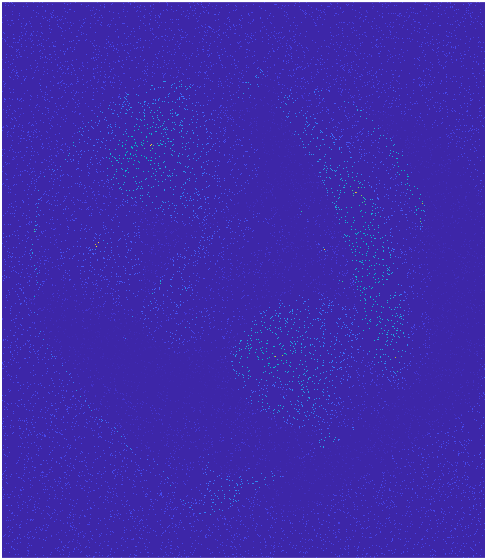} &
   \includegraphics[width=\subfigwidths]{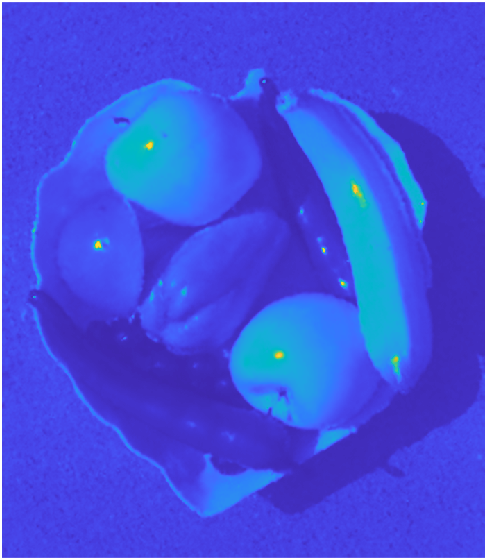} &
   \includegraphics[width=\subfigwidths]{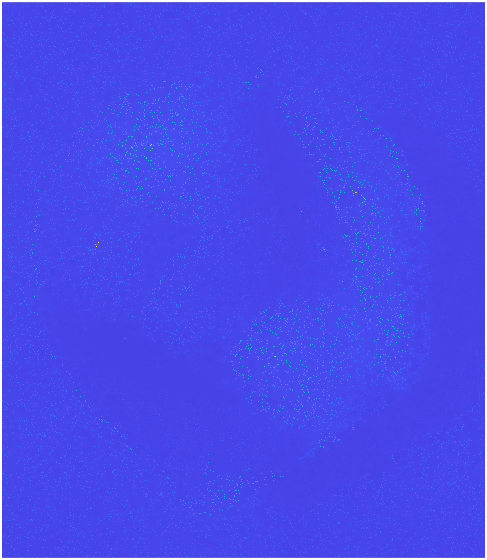} &
   \includegraphics[width=\subfigwidths]{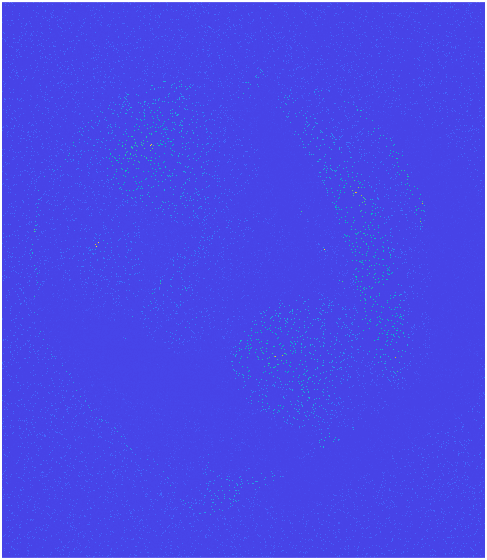} &
   \includegraphics[width=\subfigwidths]{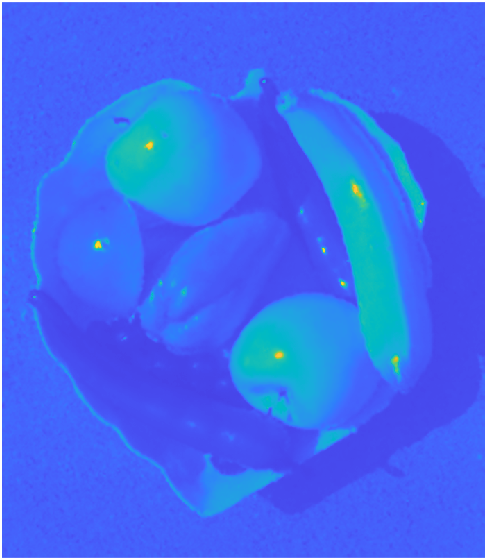} &
   \includegraphics[width=\subfigwidths]{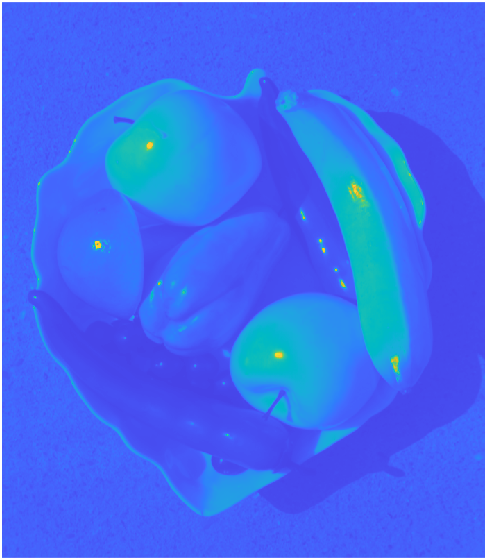} &
   \includegraphics[width=\subfigwidths]{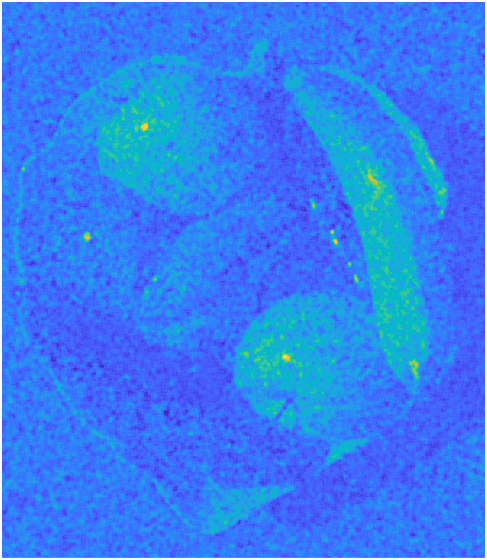} &
   \includegraphics[width=\subfigwidths]{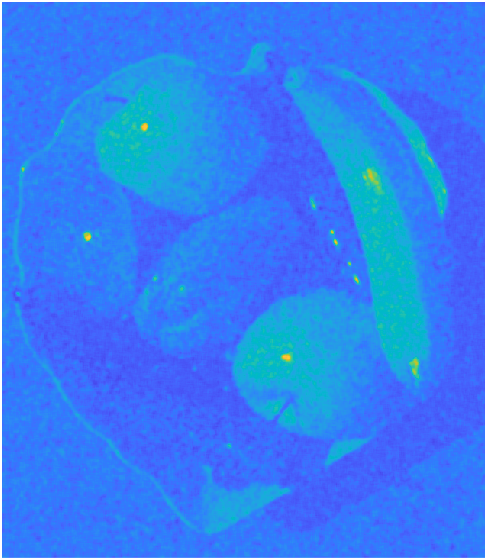} &
   \includegraphics[width=\subfigwidths]{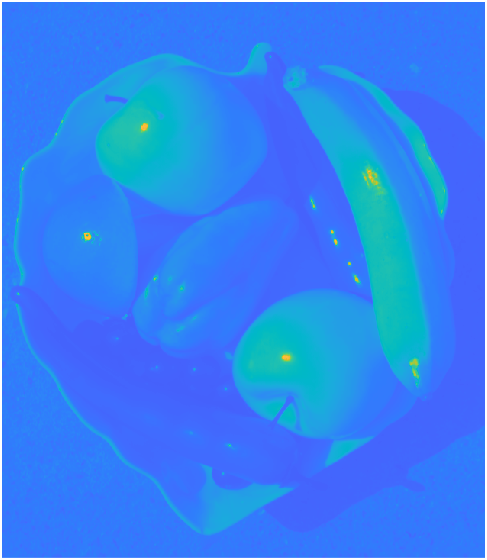} &
   \includegraphics[width=\subfigwidths]{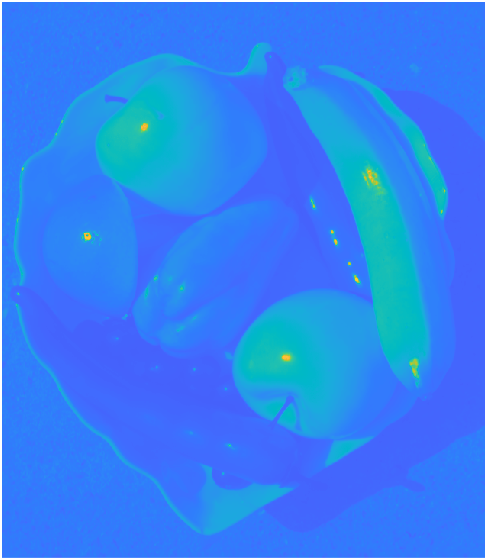}
\end{tabular}}
  \caption{Inpainting experiment with the Fru data set -- Color compositions (1st row) and 120th band (2nd row) of the images. The non-acquired (masked) pixels appear in gray in the measured images.}\label{fig_lab_map}
\end{figure*}

\begin{table}
\scriptsize
\centering
\renewcommand\arraystretch{1.5}
\caption{Inpainting experiment with the UGR data set -- Quantitative results.}\label{Tab_scenes_results}
\begin{tabular}{c|c|c|c|c|c}
\hline\hline
Methods         & PSNR $\uparrow$  & SAM $\downarrow$ & UIQI $\uparrow$ & ERGAS $\downarrow$ & SSIM $\uparrow$ \\ \hline
FastHyIn        & 38.3746          & 1.1329           & 0.8953          & 3.8298             & 0.9523          \\ \hline
Deep-HS-prior   & 35.5531          & 1.4972           & 0.8413          & 5.0447             & 0.9165          \\ \hline
WLRTR           & 38.5308          & 0.8827           & 0.916           & 3.1326             & 0.9679          \\ \hline
PnP-In         &38.6648	&1.0307	&0.9179	&4.9074	&0.9689
     \\ \hline
ADAMADMM        & 38.7166          & 0.9158           & 0.9191          & 3.3461             & 0.9700          \\ \hline
GDD         &37.1655	&1.3830	&0.8711	&4.4531	&0.9350         \\ \hline
Adam-VAE &37.6023	&1.1758	&0.9083	&4.0566	&0.9439
 \\ \hline
ADMM-VAE &38.6609	&1.1481	&0.9181	&3.2132	&0.9695
 \\ \hline
Adam-GDD  &37.6068	&1.2814	&0.8815	&4.3924	&0.9452 \\ \hline
ADMM-GDD& \textbf{39.0391} & \textbf{0.8665}  & \textbf{0.9204} & \textbf{2.9984}    & \textbf{0.9706} \\ \hline\hline
\end{tabular}
\end{table}

\begin{table}
\scriptsize
\centering
\renewcommand\arraystretch{1.5}
\caption{Inpainting experiment with the Fru data set -- Quantitative results.}\label{Tab_lab_results}
\begin{tabular}{c|c|c|c|c|c}
\hline\hline
Methods         & PSNR $\uparrow$  & SAM $\downarrow$ & UIQI $\uparrow$ & ERGAS $\downarrow$ & SSIM $\uparrow$ \\ \hline
FastHyIn        & 24.0655          & 26.3186          & 0.004           & 118.9453           & 0.2219          \\ \hline
Deep-HS-prior   & 40.7871          & 1.0734           & 0.6267          & 21.8554            & 0.9507          \\ \hline
WLRTR           & 25.2334          & 2.6928           & 0.0301          & 105.4779           & 0.4335          \\ \hline
PnP-In          & 25.6018          & 16.8758          & 0.0223          & 105.9829           & 0.4128          \\ \hline
ADAMADMM        & 40.7938          & 1.0684           & 0.6288          & 21.8406            & 0.9510          \\ \hline
GDD         &46.1837	&0.9731	&0.8011	&16.9262	&0.9868 \\ \hline
Adam-VAE &27.4481	&9.6011	&0.1758 &	96.3161	&0.5015
 \\ \hline
ADMM-VAE &32.2087	&7.0052	&0.2991	&86.6622	&0.6911
 \\ \hline
Adam-GDD       &45.9297	&0.9874	&0.8032	&17.0650	&0.9871 \\ \hline
ADMM-GDD& \textbf{46.8336} & \textbf{0.9657}  & \textbf{0.8057} & \textbf{16.6714}   & \textbf{0.9876} \\ \hline\hline
\end{tabular}
\end{table}

\subsection{Multiband image inpainting}\label{subsec:results_inpainting}
\noindent \textbf{Data --} This section reports  experiments conducted to evaluate the performance of the proposed framework when tackling a multiband image inpainting problem (see Section \ref{subsec:app_inpainting}). Two acquisition scenarios are considered and are chosen to mimic archetypal applicative contexts. These scenarios relies on two distinct data sets.

The first data set, referred to as UGR data set in what follows, is selected from the UGR Hyperspectral Image Database. It contains pairs of hyperspectral and RGB images. The hyperspectral image $\mathbf{Y}_{\mathrm{HS}}$ is of size $1000\times900$ pixels with $B=61$ spectral bands. It was acquired using the hyperspectral camera V-EOS by Photon etc and captured outdoor environments in Granada, Spain. The associated RGB image $\mathbf{Y}_{\mathrm{HR}}$ is of size $1000\times900\times3$ and is rendered by CIE Standard Illuminant D65 with the CIE 1931 $2^{\circ}$ Standard Observer with gamma correction ($\gamma = 0.6$). To mimic a sensor default, the masks $\boldsymbol{\Omega}_{\mathrm{b}}$ and $\boldsymbol{\Omega}_{\mathrm{p}}$ are defined such that $25$ randomly selected bands are contaminated by stripe-like noise, i.e., with dead pixels on randomly selected columns. A color composition of the image and a given band are depicted in Fig.~\ref{fig_scene2_map} (1st column).

The second data set, termed as Fru data set, was acquired in our laboratory by a GaiaField camera. The image $\mathbf{Y}_{\mathrm{HS}}$ is of size $696\times 801$ pixels with $256$ spectral bands covering a spectral range from $400$nm to $1000$nm, with spectral resolution up to $0.58$nm. An auxiliary RGB camera in the hyperspectral device was utilized to acquire the corresponding high resolution image $\mathbf{Y}_{\mathrm{HR}}$. To mimic a spatial partial acquisition, the bandwise mask $\boldsymbol{\Omega}_{\mathrm{b}}$ is chosen as $\boldsymbol{\Omega}_{\mathrm{b}}=\mathbf{I}_{B}$ ($B=\tilde{B}$) and the pixelwise mask $\boldsymbol{\Omega}_{\mathrm{p}}$ is chosen to randomly select $5\%$ of the spatial locations. In other words, only $5\%$ of the spectra in $\mathbf{Y}_{\mathrm{HR}}$ are available to reconstruct the full image.  The reference image and the corresponding measurements are shown in the first and second subimages of Fig.~\ref{fig_lab_map}.\\

\noindent \textbf{Compared methods --} To evaluate the efficiency of the proposed framework, it is compared to several state-of-the-art algorithms, namely
FastHyIn~\cite{zhuang2018fast}, Deep-HS-prior~\cite{sidorov2019deep}, weighted low-rank tensor recovery(WLRTR)~\cite{chang2020weighted}, PnP-In~\cite{lai2022deep}, ADAM-ADMM~\cite{lin2021admm} and GDD~\cite{uezato2020guided} as compared methods. The FastHyIn is a fast and competitive inpainting algorithm based on low-rank and sparse representations. The Deep-HS-prior is based on a deep image prior framework. The WLRTR is a unified low-rank tensor recovery method. The PnP-In is a recently proposed plug-and-play based inpainting method, which can plug denoiser priors for which we set $\rho=1\times 10^{-4}$ and directly use the pretrained Gray/RGB denoisers FFDNet~\cite{zhang2018ffdnet} as the denoiser.
The ADAM-ADMM solves the inpainting problem and integrates deep prior information. We use the restored results of Deep-HS-prior as the deep prior regularization with $\mu_a$ set to $1\times 10^{-3}$ and $\lambda_a$ set to $0.01$. Adam-VAE and Adam-GDD are also used as compared methods. To pretrain deep models embedded into the proposed framework, the number of epochs and the learning rate are set to $100$ and $1\times 10^{-3}$ for the VAE and to $5000$ and $0.01$ for the  GDD.  As in the previous experiment, the VAE has been trained using 3000 patches from the guidance image and 10000 patches from an additional image data set, with patch size chosen as  $25\times 25$. The parameters $\mu$ and $\lambda$ are set to $1\times 10^{-3}$ and $1\times 10^{-5}$.\\

\noindent \textbf{Results --} The quantitative figures-of-merit PSNR, SAM, UIQI, ERGAS and SSIM obtained by the compared algorithms on the UGR-data set are reported in Table~\ref{Tab_scenes_results}. All compared methods provide satisfactory reconstructed results. The proposed framework instantiated with an ADMM optimization scheme and a GDD  regularization outperforms all compared methods. The enhancement stems from two aspects: the first one comes from the guided prior learnt from the complementary RGB image, the other is the integrating use of convex optimization and deep generative priors. Fig.~\ref{fig_scene2_map} shows the inpainting results for this data set. It is clear that the proposed method is  closer to the ground-truth image. Table~\ref{Tab_lab_results} reports the quantitative results and Fig.~\ref{fig_lab_map} depicts the reconstructed images obtained on the Fru data set. For this challenging task, for which  some spectra  are completely unavailable for a large part of the spatial positions, conventional hyperspectral inpainting algorithms, such as FastHyIn, WLRTR and PnP-In, fail to recover the missing pixels and get bad restored results. Whereas, deep learning-based inpainting methods or approaches integrating deep networks provide good results. In particular the proposed method ADMM-GDD achieves the best results. It is worth noting that the result of ADMM-VAE is still blurry and noisy, which may due to the limited generative ability of the VAE model which does not fully exploit the guidance image $\mathbf{Y}_{\mathrm{HR}}$.

\begin{figure*}[t]
  \centering
  \includegraphics[width=18cm]{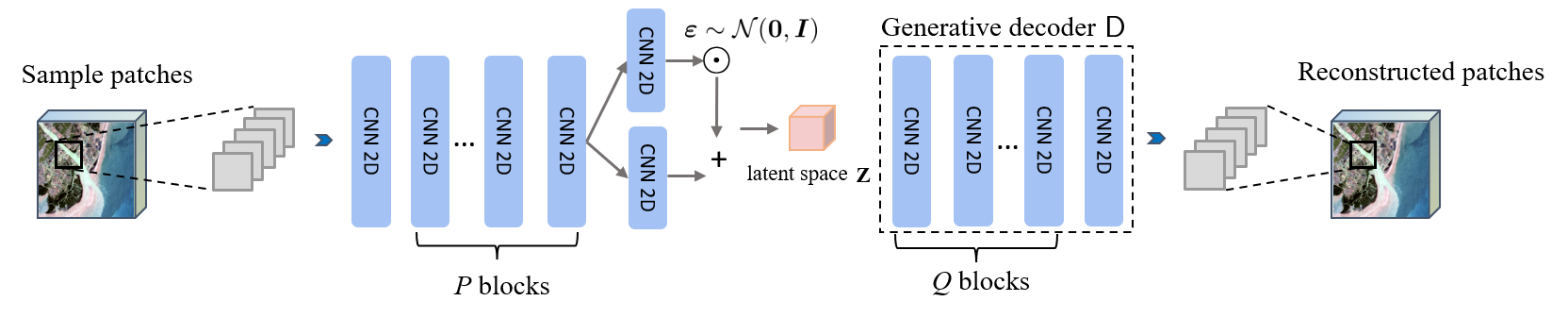}\\
  \caption{The architecture of the VAE model.}\label{fig_frame_network_VAE}
\end{figure*}

\section{Conclusion}
\label{sec:conclusion}
This paper  proposed a generic framework for multiband imaging inverse problems. It relies on a  guided deep regularization  designed to embed a generative prior learnt from an auxiliary acquisition of high spatial resolution. The resulting nonlinear objective function was minimized using an alternating direction method of multipliers. Contrary to a brute force method, i.e., based on automatic differentiation (e.g., Adam), this splitting based strategy had the great advantage of decomposing the initial problem into three simpler subproblems. In particular, for most of the inverse problems of interest, it was shown that one could resort to closed-form expressions of one subproblem. A nonlinear optimizer could be used for minimization the subproblem involving the deep regularization, which amounted to perform a nonlinear projection onto the range of the generative model. As a particular instance, the generative model was chosen as a guided deep decoder. However the proposed framework appeared to be sufficiently flexible to let the choice of the deep regularization to the end-user. The proposed framework was instantiated for two particular yet ubiquitous multiband imaging tasks, namely fusion and inpainting. Experiments conducted for these two tasks showed that the proposed framework outperformed state-of-the-art algorithms. Future work may include  unrolling (or unfolding) the derived iterative optimization procedure to jointly learn the hyperparameters.

\appendix[VAE-based generative model]

VAE has demonstrated excellent performance to model probability distributions of complex data sets and to generate new samples similar to the observations \cite{kingma2013auto}. In this work, a VAE is used as a generative model and an isotropic Gaussian prior is typically imposed on the latent vectors $\mathbf{Z}$.
As illustrated in Fig. \ref{fig_frame_network_VAE}, the architecture of this network consists of an encoder and a decoder. The encoder maps input image patches into the latent feature space $\mathcal{Z}$, and the decoder is trained to reconstruct the input image patches. The core of the network exploits convolutional layers. Apart from the input and output layers, the encoder and decoder consist of $P$ and $Q$ blocks, respectively. Each block is composed of a $3\times 3$ convolution layer and a LeakyReLU activation function. In our work, the $2$-dimensional input image patches used to train the network are randomly selected from the guidance image and complementary images available from widely used image data sets such as DOTA~\cite{xia2018dota} and HRSC2016~\cite{su2020hq}.
The objective function is then defined as
\begin{equation}\label{eq.loss_VAE}
\begin{aligned}
\mathcal{L}_\text{VAE} &=\frac{1}{N} \sum_{n=1}^{N} \left\|\widehat{\mathbf{x}}_{n}-\mathbf{x}_{n}\right\|_{\text{F}}^{2} \\
&+\lambda_{\mathrm{K L}} \frac{1}{N} \sum_{n=1}^{N} D_{\mathrm{K L}}\left(q\left(\boldsymbol{z}_{n} \mid \mathbf{x}_{n}\right) \| p(\boldsymbol{z})\right),\\
\end{aligned}
\end{equation}
where $\mathbf{x}_{n}$ and $\widehat{\mathbf{x}}_{n}$ denote the $n$th input and reconstructed patches, respectively. In \eqref{eq.loss_VAE}, $\lambda_{\mathrm{K L}}$ is a hyperparameter adjusting the weight between the reconstruction error and the Kullback-Leibler (KL) divergence between the posterior and instrumental distributions. This network is trained thanks to the Adam optimizer. After training, the decoder is used as a generative model $\mathsf{D}(\cdot)$ which embeds the spatial regularization. 

\bibliographystyle{IEEEtran}
\bibliography{IEEEfull,BIB}\ 

\end{document}